\documentclass[journal]{new-aiaa}
\usepackage[utf8]{inputenc}
\usepackage{textcomp}

\usepackage{graphicx}
\usepackage{amsmath}
\usepackage[version=4]{mhchem}
\usepackage{siunitx}
\usepackage{longtable,tabularx}
\title{
Orbital Facility Location Problem for Satellite Constellation Servicing Depots
}

\author{
Yuri Shimane
\footnote{Ph.D. Student, Daniel Guggenheim School of Aerospace Engineering.
Student Member AIAA},
Nicholas Gollins
\footnote{Ph.D. Student, Daniel Guggenheim School of Aerospace Engineering.
Student Member AIAA}, and
Koki Ho
\footnote{Associate Professor, Daniel Guggenheim School of Aerospace Engineering. Senior Member AIAA.}
}
\affil{Georgia Institute of Technology, Atlanta, Georgia 30332}

\usepackage{amsmath}
\usepackage{todonotes}
\usepackage{siunitx}
\usepackage{booktabs}
\usepackage{mathtools}
\usepackage{caption}
\usepackage{subcaption}
\usepackage{multirow}
\usepackage{lscape} 
\newcommand{\elements}{\text{\oe}}
\newcommand{\mddry}{m_{d,\text{dry}}}
\newcommand{\msdry}{m_{s,\text{dry}}}
\newcommand{\mdwet}{m_{d,\text{wet}}}

\newcommand{\msL}{m_{s,L}}
\newcommand{\mlmax}{m_{l,\max}}

\begin{document}

\maketitle

\begin{abstract}
This work proposes an adaptation of the Facility Location Problem for the optimal placement of on-orbit servicing depots for satellite constellations in high-altitude orbit. The high-altitude regime, such as Medium Earth Orbit (MEO), is a unique dynamical environment where low-thrust propulsion systems can provide the necessary thrust to conduct plane-change maneuvers between the various orbital planes of the constellation. 
As such, on-orbit servicing architectures involving servicer spacecraft that conduct round-trips between servicing depots and the client satellites of the constellation may be conceived. 
To this end, a new orbital facility location problem formulation is proposed based on binary linear programming, in which the costs of operating and allocating the facility(ies) to satellites are optimized in terms of the sum of the Equivalent Mass to Low Earth Orbit (EMLEO).
The low-thrust transfers between the facilities and the clients are computed using a parallel implementation of a Lyapunov feedback controller. 
The total launch cost of the depot along with its servicers, propellant, and payload are taken into account as the cost to establish a given depot. 
The proposed approach is applied to designing an on-orbit servicing depots architecture for the Galileo and the GPS constellations. 
\end{abstract}

\section*{Nomenclature}


{\renewcommand\arraystretch{1.0}
\noindent\begin{longtable*}{@{}l @{\quad=\quad} l@{}}
$a$        & semimajor axis \\
$\boldsymbol{c}_{ij}$   & facility allocation cost \\
$D$        & demand of servicing trips \\
$e$        & eccentricity  \\
$\boldsymbol{f}_j$      & facility usage cost \\
$g_0$      & standard acceleration due to gravity \\
$i$        & inclination \\
$i_{\text{sp}}$ & specific impulse \\
$m$        & number of clients \\
$\mddry$   & depot dry mass \\
$\mlmax$   & maximum launch mass of launch vehicle \\
$\msdry$   & servicer dry mass \\
$n$        & number of candidate facility sites \\
$\msL$     & servicer payload mass \\
$r_{a}$    & apogee radius \\
$r_{p}$    & perigee radius \\
$\boldsymbol{x} $    & spacecraft orbital elements \\
$X_{ij}$   & binary allocation variable \\
$Y_j$      & binary usage variable \\
$\theta$      & true anomaly \\
$\mu$      & gravitational parameter \\
$\phi$        & initial-to-final mass-ratio \\
$\Omega$   & right-ascension of ascending node \\
$\omega$   & argument of periapsis 
\end{longtable*}}

\section{Introduction}
On-orbit servicing, assembly, and manufacturing (OSAM) is a key piece of technology gaining attention from both commercial and governmental players alike \cite{nasa_osam_report,aerospace_osam,duke_osam}. 
Examples include the Robotic Servicing of Geosynchronous Satellites (RSGS) program by the Defense Advanced Research Projects Agency (DARPA) \cite{Sullivan2018} and Lockeed Martin's subsidiary SpaceLogistics \cite{NorthropSpaceLogistics} for servicing geostationary orbit (GEO) satellites, or NASA's OSAM-1 \cite{Reed2016} that focuses on polar low-Earth orbit (LEO). 
The primary motivation for OSAM has been the cost reduction that comes with extending the lifespan of satellites \cite{Saleh2002,Lamassoure2002,Saleh2003,Long2007}. 
It is also possible to recognize the possibility for OSAM to contribute to a more sustainable practice in the space sector, both in terms of the orbital environment and the environmental impact of rocket launch \cite{Sirieys2022}. 
Conceptualized OSAM needs span multiple activities, including refueling, spearheaded by private entities such as Orbit Fab \cite{OrbitFabHydrazine} and Astroscale U.S. Inc. \cite{AstroscaleUSNews}, repair/refurbishment/mission extension by retrofitting additional modules \cite{Hall1999}, inspection, assembly, or end of life (EOL) services, typically consisting of de-orbiting or tugging to graveyard orbits \cite{nasa_osam_report,aerospace_osam}. 
A detailed review of the robotics technology that will enable future OSAM activities is provided by Flores-Abad et al \cite{FloresAbad2014} and more recently by Moghaddam \cite{Moghaddam2021}. 

To date, multiple studies on OSAM applications for GEO and low-inclination geosynchronous orbit (GSO) satellites have been conducted, driven by the relatively high cost of assets lying in these orbits \cite{Galabova2006,Hudson2020,Sarton2021a,Sarton2021b}. A particular convenience of GEO servicing comes from the fact that all potential client satellites lie on the equatorial plane; hence, the servicer simply needs to conduct a phasing maneuver to deliver the service. 
Sarton du Jonchay and Ho \cite{SartonduJonchay2017b} studied servicing architectures including both MEO and GEO satellites, but assumed these to be coplanar. 
Meanwhile, studies for LEO has also been considered by Luu and Hastings \cite{Luu2020,Luu2021,Luu2022} and Sirieys et al \cite{Sirieys2021}. Luu and Hastings \cite{Luu2021} also provides a review of ongoing work on OSAM for LEO constellations. Converse to GEO constellations, LEO constellations exist across multiple orbital planes, but strong $J_2$ perturbations may be leveraged to use differential drift to the servicer's advantage. 

A substantially different scenario from an astrodynamics perspective arises when servicers must navigate between orbital planes of the constellation by primarily utilizing their own propulsion system. Such maneuvers are commonly prohibitively expensive for chemical thrusters, thus requiring the use of low-thrust propulsion with high specific impulse. 
This scenario is relevant when the constellation to be serviced is at high altitudes, such as Medium Earth Orbit (MEO) or high-inclination GSO. 
Fortunately, if the servicing is limited to a single constellation, certain features such as the semimajor axis, inclination, eccentricity, and argument of perigee are commonly shared among the constellation fleet. 
Furthermore, OSAM needs such as refueling, retrofitting, and orbit alteration, which are the primary OSAM needs that have recently occurred from both private and governmental players \cite{Astroscale,spacenews_orbitfab_hydrazine,diu_news,maxar}, can be planned well in advance. This gives the servicing vehicle enough time to carry out an economical transfer with a longer duration to reach its clients. 
Hall and Papadopoulous \cite{Hall1999} have previously reported on the hardware considerations, while Leisman et al \cite{Leisman1999} conducted a systems engineering study for servicing the GPS constellation. 

The placement of facilities is a fundamental problem in terrestrial logistics. The facility location problem (FLP), since its initial introduction by Cooper \cite{Cooper1963}, has been studied extensively over the past decades \cite{Hakimi1965,Cornuejols1977,Rosenwein1994DiscreteLT,SimchiLevi2005,flp_WOLF2011}. 
There exist many flavors of FLPs, notably with the distinction between discrete and continuous formulations. 
In the discrete case, a discrete set of potential locations are considered, with a flexible number of facilities. In contrast, in the continuous case, facility locations are considered in continuous space, but the number of facilities must typically be prescribed \cite{laporte2015location}. 

The FLP has seen a wide range of applications, including emergency services \cite{Matsutomi1992}, telecommunications \cite{Fortz2015}, healthcare \cite{AhmadiJavid2017}, waste disposal management \cite{Franco2021}, and disaster response \cite{Stienen_Wagenaar_denHertog_Fleuren_2021}. 
In the context of space applications, McKendree and Hall \cite{McKendree2005} studied a single-facility variant of the FLP to locate an interplanetary manufacturing plant. 
Dorrington and Olsen \cite{Dorrington2019} applied a variant of the FLP to the asteroid mining problem. 
Finally, Zhu et al \cite{Zhu2020} apply the FLP for OSAM depots in Sun-synchronous orbits (SSO). 

This work proposes a method to identify optimal combinations of facility locations where the servicing depots are to be placed for an OSAM architecture of a particular satellite constellation. 
This is done by adopting the FLP to the on-orbit servicing depot of high-altitude satellite constellations; the proposed formulation is coined as the Orbital Facility Location Problem (OFLP). 
The OFLP is able to simultaneously find the optimal number of facilities, their orbital position within a discretized set of candidate slots, as well as the optimal allocation of each client to the appropriate facility. 
The formulation is agnostic to the specific type of servicing and is compatible with any service that involves a mass drop-off to the client satellite, such as on-orbit refueling \cite{VANDENKERCKHOVE1982} or retrofitting of a module \cite{Hall1999}.

The two key modifications of the OFLP come from (1) the formulation of the allocation cost, and (2) the formulation of a facility's usage cost. 
As is the case in terrestrial logistics, the optimal location(s) to place facilities depend on some form of a ``distance'' metric between the facility and the client(s) it must service. Furthermore, the use of an additional facility has associated incurred costs, which must also be taken into account. 
However, in the context of space-based facilities, the ``distance'' cannot be simple Euclidean norms or Manhattan distance, but must rather account for the cost of the trajectory to be taken by the servicer. 
In this work, a Lyapunov feedback controller known as Q-Law \cite{Petropoulos2003,Petropoulos2004,Petropoulos2005} is utilized to compute the cost associated with conducting a return trip between a facility and its client satellite using low-thrust propulsion. The total propellant required for these two maneuvers is used as the ``distance''. 
To the best of the authors' knowledge, this work is the first attempt at considering low-thrust transfers in an FLP framework. 

The costs incurred in building a facility at a given location in space are considered on the basis of the launch and orbit insertion cost of the depot carrying with it its servicers, propellant, and payload. 
In order to coherently consider the orbit transfer costs between the facilities and clients as well as the facilities' building cost through a single objective function, this work formulates the optimization problem in terms of effective mass to LEO (EMLEO). 

The proposed OFLP framework enables the non-trivial, optimal placement of depots for OSAM of spatially distributed clients such as for satellite constellations. 
By leveraging the discrete version of the FLP, the number of facilities does not need to be predetermined, thus allowing parametric studies in terms of high-level parameters of the architecture, such as the dry mass of the depot or the number of trips to be made to each client. 

Once an architecture design is obtained by the OFLP, this work further proposes a refinement scheme whereby the location of each facility is re-optimized in continuous space, while freezing the allocations determined by OFLP. This increases the fidelity of the designed OSAM architecture as the facility location is no longer restricted by the coarseness of the discretized orbital slots. 

This paper is organized as follows. In Section \ref{sec:Astrodynamics}, the design of low-thrust transfers and the associated approach for estimating the rendez-vous cost for a given pair of orbital elements is introduced. 
Then, Section \ref{sec:Methods} discusses the FLP along with its adaptation to the OSAM case. 
Section \ref{sec:numerical_results} presents numerical results applying the formulation to service a combination of the GPS and Galileo constellations. 
Finally, Section \ref{sec:conclusion} provides a conclusion to this work.

\section{Astrodynamics Background}
\label{sec:Astrodynamics}
Typical on-orbit servicing demands are well-known in advance; this results in phasing maneuvers having little impact on overall transfer costs, as arbitrarily long phasing maneuver time may be utilized to minimize propellant expenditure. 
As such, when considering the transfer cost in the on-orbit servicing context, the primary propellant expenditure comes from the orbit transfer cost, i.e. the cost associated with modifying the orientation and shape of the orbit in 3D space. 
To this end, a Lyapunov feedback controller for spacecraft trajectories, commonly referred to in the astrodynamics literature as Q-law, is employed. 

\subsection{Orbit Transfer via Q-Law}
The orbital transfer is conducted using Q-law, a Lyapunov-function-based feedback control law first introduced by Petropoulos \cite{Petropoulos2003,Petropoulos2004,Petropoulos2005}. 
Despite its sub-optimal nature, its ability to rapidly compute multi-revolution low-thrust transfers has made it a popular tool, particularly for large-scale preliminary transfer designs \cite{Jagannatha2019,Epenoy2019,Shannon2020,Holt2021,Wijayatunga2021}. 
The feedback law consists of determining the optimal thrust direction given the current and target orbital elements. Among the target elements, only the five slow variables are used for the transfer, as the considered problem does not necessitate rendez-vous. 

The dynamics of the satellite may be expressed in terms of orbital elements via Gauss' variational equations, or the Variation of Parameters (VOP) equations. Let the state be given by $\boldsymbol{x} \in \mathbb{R}^6$, and consider $\boldsymbol{B} \in \mathbb{R}^{6 \times 3}$ and $\boldsymbol{D} \in \mathbb{R}^6$, such that
\begin{equation}
    \dot{\boldsymbol{x}} = \boldsymbol{B}(\boldsymbol{x})\boldsymbol{F} + \boldsymbol{D} (\boldsymbol{x})
    \label{eq:vop}
\end{equation}
where $\boldsymbol{F} \in \mathbb{R}^3$ is the perturbing force expressed in terms of radial, tangential, and normal components
\begin{equation}
    \boldsymbol{F} = [F_r, F_{\theta}, F_n]^{\mathrm{T}}
\end{equation}
The set of elements $\boldsymbol{x}$ may be Keplerian elements $[a,e,i,\Omega,\omega,\theta]$ or other alternative sets of elements.
While the original Q-law has been developed in terms of Keplerian elements, the well-known singularities at $i=0$ and $e=0$ are problematic as these are typical orbital elements in which a spacecraft may be placed. For this reason, the use of alternative element sets, such as the modified equinoctial elements (MEE), given in terms of Keplerian elements as 
\begin{equation}
    \begin{aligned}
        & p = a(1-e^2)   \\
        & f = e \cos{(\Omega + \omega)}  \\
        & g = e \sin{(\Omega + \omega)}\\
        & h = \tan{\left(\frac{i}{2}\right)} \cos \Omega  \\
        & k = \tan{\left(\frac{i}{2}\right)} \sin \Omega  \\
        & L = \Omega + \omega + \theta
    \end{aligned}
\end{equation}
is beneficial as the singularity is moved to $i = \pi$. 
Furthermore, the use of the MEE with the semi-parameter $p$ replaced by the semimajor axis $a$ has been previously reported to yield convergence benefits \cite{Varga2016}, and is employed in this work as well. 
For this set of elements $\boldsymbol{x} = [a,f,g,h,k,L]$, the VOP are given by
\begin{equation}
    \boldsymbol{B}(\boldsymbol{x}) = \left[\begin{array}{ccc}
        \dfrac{2 a^{2}}{h} e \sin \theta & \dfrac{2 a^{2}p}{hr} & 0 
        \\[0.1em]
        \sqrt{\dfrac{p}{\mu}} \sin L & \sqrt{\dfrac{p}{\mu}} \dfrac{1}{w}[(w+1) \cos L+f] & -\sqrt{\dfrac{p}{\mu}} \dfrac{g}{w}[h \sin L-k \cos L] 
        \\[0.1em]
        -\sqrt{\dfrac{p}{\mu}} \cos L & \sqrt{\dfrac{p}{\mu}} \dfrac{1}{w}[(w+1) \sin L+g] & \sqrt{\dfrac{p}{\mu}} \dfrac{f}{w}[h \sin L-k \cos L] 
        \\[0.1em]
        0 & 0 & \sqrt{\dfrac{p}{\mu}} \dfrac{s^{2}}{2 w}  \cos L
        \\[0.1em]
        0  & 0 & \sqrt{\dfrac{p}{\mu}} \dfrac{s^{2}}{2 w} \sin L
        \\[0.1em]
        0 & 0  & \sqrt{\dfrac{p}{\mu}} \dfrac{1}{w} [h \sin L-k \cos L]
    \end{array}\right]
    \label{eq:vop_Amatrix}
\end{equation}
and
\begin{equation}
\boldsymbol{D}(\boldsymbol{x}) =\left[\begin{array}{llllll}
0 & 0 & 0 & 0 & 0 & \sqrt{\mu p}\left(\dfrac{w}{p}\right)^{2}
\end{array}\right]^{\mathrm{T}}
\end{equation}
\begin{align}
    w &= 1 + f \cos L + g \sin L
    \\
    s^2 &= 1 + h^2 + k^2
\end{align}
Note that $\boldsymbol{F}$ can be due to any form of perturbing acceleration, such as propulsive force, atmospheric drag, third-body attraction, or non-spherical gravity. 
The acceleration due to the propulsive force, which is to be determined to guide the spacecraft to its target orbit, is given by
\begin{equation}
    \boldsymbol{F} = \dfrac{T}{m}\left[
        \cos \alpha \cos \beta , \,
        \sin \alpha \cos \beta , \,
        \sin \beta
    \right]^{\mathrm{T}}
    \label{eq:F_vector_eqn}
\end{equation}
where $\alpha$ and $\beta$ are the in-plane and out-of-plane angles, respectively. 

Denoting the osculating element $\elements \in [a,f,g,h,k]$ and the targeted elements as $\text{\oe}_T$, the Lyapunov function is given by
\begin{equation}
    Q = 
    \left(1+W_{p} P\right) \sum_{\text{\oe}}
    S_{\text{\oe}} W_{\text{\oe}}
    \left(
        \dfrac{\text{\oe} - \text{\oe}_T}{\dot{\text{\oe}}_{x x}}
    \right)^{2}
    \, , 
    \quad \text{\oe} = a, f, g, h, k
    \label{eq:lyap_function}
\end{equation}
In essence, $Q$ penalizes the difference between $\elements$ and $\elements_{T}$ through the subtraction term in the summation. 
$S_{\elements}$ is a scaling factor given by
\begin{equation}
    S_{\elements} 
    =
    \begin{cases}
        \left[1+\left(\dfrac{\left|a-a_{T}\right|}{\sigma a_{T}}\right)^{\nu}\right]^{1 / \zeta},
        & \elements = a
        \\
        1, & \text{ otherwise }
    \end{cases}
\end{equation}
where $\sigma$, $\nu$ and $\zeta$ are scalar coefficients, which prevents non-convergence of $a \to \infty$. This is necessary as when $a \to \infty$, the $\dot{\elements}_{x x}$ terms also tend to $\infty$ and thus $Q$ is reduced, however, this is not a physically useful solution. $W_{\elements}$ are scalar weights that may be assigned to different elements, if one is to be favored for targeting over another. $P$ is a penalty term on the periapsis radius, given by
\begin{equation}
    P =\exp \left[k_{r_p}\left(1-\frac{r_{p}}{r_{p \min }}\right)\right]
\end{equation}
where $r_p$ is the current orbit's periapsis radius given by 
\begin{equation}
    r_p = a (1 - e)
\end{equation}
and $r_{p \min }$ is a user-defined threshold. Here, $k_{r_p}$ is also a pre-defined constant on this penalty term and represents the gradient of this exponential barrier function near $r_{p \min }$. $W_p$ is a scalar weight to be placed on the periapsis penalty term. 
The $\dot{\elements}_{x x}$ terms represent the maximum rates of change of a given orbital element with respect to the thrust direction and true anomaly along the osculating orbit and are given in the Appendix.

Through the application of Lyapunov control theory, the Q-law strategy consists of choosing the control angles $\alpha$ and $\beta$ such that the time-rate of change of $Q$ is minimized at each time-step
\begin{equation}
    \min_{\alpha, \beta} \dot{Q}
    \label{eq:minQ_def}
\end{equation}
where $\dot{Q}$ can be expressed using the chain rule as 
\begin{equation}
    \begin{aligned}
        \dot{Q} &= \sum_{\elements } \dfrac{\partial Q}{\partial \elements} \dot{\elements}
        = D_1 \cos \beta \cos \alpha 
        + D_2 \cos \beta \sin \alpha 
        + D_3 \sin \beta
        \,, \quad \elements=a, f, g, h, k
    \end{aligned}
    \label{eq:qdot}
\end{equation}
where
\begin{equation}
    \begin{aligned}
        D_1 &= \sum_{\elements} \dfrac{\partial Q}{\partial \elements} \dfrac{\partial \dot{\elements}}{\partial F_{\theta}}
        \\
        D_2 &= \sum_{\elements} \dfrac{\partial Q}{\partial \elements} \dfrac{\partial \dot{\elements}}{\partial F_{r}}
        \\
        D_3 &= \sum_{\elements} \dfrac{\partial Q}{\partial \elements} \dfrac{\partial \dot{\elements}}{\partial F_{n}}
    \end{aligned}
\end{equation}
The choice of $\alpha$ and $\beta$ based on condition \eqref{eq:minQ_def}, given by
\begin{equation}
    \begin{aligned}
        \alpha^* &= \arctan(-D_2, -D_1)
        \\
        \beta^* &= \arctan\left( \dfrac{-D_3}{\sqrt{D_1^2 + D_2^2}} \right)
    \end{aligned}
\end{equation}
ensures the fastest possible decrease of $Q$, thereby providing the best immediate action for the spacecraft to take to arrive at $\elements_T$. 
Note that while $\dot{\elements}$ consists simply of the first 5 rows of the VOP given in expression \eqref{eq:vop_Amatrix}, the expression for $\frac{\partial Q}{\partial \elements}$ is cumbersome to derive analytically. Instead, a symbolic toolbox is used to obtain these expressions. 
%
Finally, it is noted that there are several work in the past that extended the Q-law scheme to include a mechanism for coasting \cite{Petropoulos2005,Holt2021} as well as other perturbing forces such as oblateness or three-body effects \cite{Varga2016,Jagannatha2019,Shannon2020}. 
While these variants are not considered in this work, they may be used in place of the basic controller in the OFLP, which will be presented subsequently.

\subsection{Launch and Insertion Costs of Depot}
The launch and insertion cost of a given depot depends on the mass of the depot and its servicer(s) as well as its orbital elements. In the context of constellation servicing, the depots are to be placed in an orbit sharing the same inclination as the constellation fleet. Hence, out of the six Keplerian elements, only the semimajor axis and eccentricity have an effect on the launch and insertion cost. 

In order to place the depot into its desired orbital slot, the launch vehicle must first lift off the depot from the launch pad and place it into a transfer orbit, and the depot must then conduct an insertion maneuver upon arrival at its desired orbit. 
Both the insertion cost into the transfer orbit and the insertion cost to the depot's desired orbit depend on the choice of the orbital slot; as such, this work considers the effective mass to LEO (EMLEO) as the standardized metric for evaluating the depot's equivalent, total insertion cost as a standardized mass. 
EMLEO is the hypothetical mass the depot would have if it were to first be placed into a circular LEO on the same orbital plane as the transfer orbit, rather than into the transfer orbit itself, by the launch vehicle. 
It thus gives a consistent cost to compare depots that have differing transfer orbit insertion costs. 

In this work, we assume a coplanar Hohmann transfer to deliver a depot. With the aforementioned definition, the EMLEO of a given orbital slot can be found by calculating the propellant mass needed to perform a Hohmann transfer from a (hypothetical) LEO with radius $r_0$ to the desired orbital slot. 
The first impulse of the Hohmann transfer corresponds to the transfer orbit insertion, which is (hypothetically) carried out by the launch vehicle, while the second impulse of the Hohmann transfer is carried out by the depot itself. 
Using the rocket equation, the combined mass-ratio $\phi$ is the product of the mass-ratio due to the first burn by the launch vehicle $\phi^{(l)}$ and the mass-ratio due to the second burn by the depot itself $\phi^{(d)}$, and can be expressed as 
\begin{equation}
    \phi = \phi^{(d)} \phi^{(l)} =  \exp{ \left( \dfrac{\Delta V_2}{g_0 I_{\text{sp},d}} \right)} 
    \exp{ \left( \dfrac{\Delta V_1}{g_0 I_{\text{sp},l}} \right)}
    \label{eq:massratio_launch_insertion}
\end{equation}
where $\Delta V_1$ and $\Delta V_2$ are the impulsive burn magnitudes of the two maneuvers, $g_0$ is the standard acceleration due to gravity, $I_{\text{sp},d}$ is the specific impulse of the depot, and $I_{\text{sp},l}$ is the launch vehicle. 
To compute the $\Delta V$ magnitudes, the perigee, and apogee are first computed from the Keplerian elements of the $j^{\mathrm{th}}$ depot via
\begin{equation}
    r_p = a(1-e) \,,\,\,\, r_a = a (1+e)
\end{equation}
To transfer from a circular orbit to an elliptical orbit via a Hohmann transfer, the elliptical transfer orbit would have a periapsis at the radius of the initial circular orbit and an apoapsis at either the final orbit's perigee or apogee. Since $I_{\text{sp},d}$ and $I_{\text{sp},l}$ are not necessarily the same, both options are considered, and the strategy resulting in a smaller mass-ratio for the $j^{\mathrm{th}}$ depot, $\phi_j$, is chosen. 
In each case, the $\Delta V$'s are given by
\begin{equation}
    \begin{cases}
        \begin{aligned} 
            \Delta V_{p,1} &= \sqrt{\mu\left(\dfrac{2}{r_0} - \dfrac{2}{r_0 + r_p}\right)} - \sqrt{\dfrac{\mu}{r_0}}
            \\
            \Delta V_{p,2} &= \sqrt{\mu\left(\dfrac{2}{r_p} - \dfrac{1}{a}\right)} 
            - \sqrt{\mu\left(\dfrac{2}{r_p} - \dfrac{2}{r_0 + r_p}\right)}
        \end{aligned} & \text{ if second burn at }r_p
    \\[4.0em]
        \begin{aligned} 
            \Delta V_{a,1} &= \sqrt{\mu\left(\dfrac{2}{r_0} - \dfrac{2}{r_0 + r_a}\right)} - \sqrt{\dfrac{\mu}{r_0}}
            \\
            \Delta V_{a,2} &= \sqrt{\mu\left(\dfrac{2}{r_a} - \dfrac{1}{a}\right)} 
            - \sqrt{\mu\left(\dfrac{2}{r_a} - \dfrac{2}{r_0 + r_a}\right)}
        \end{aligned} & \text{ if second burn at }r_a
    \end{cases}
\end{equation}
Then, the mass-ratio for $\phi_j$ is the combination of $\phi^{(d)}$ and $\phi^{(l)}$ that results in a smaller value 
\begin{equation}
    \phi_j = \min \left\{ \phi_{p,j}, \phi_{a,j} \right\}
    \label{eq:massratio_j_launch_insertion}
\end{equation}
where
\begin{equation}
    \begin{aligned}
        \phi_{p,j} = (\phi^{(d)} \phi^{(l)})_p &= \exp{ \left( \dfrac{\Delta V_{p,2}}{g_0 I_{\text{sp},d}} \right)} 
        \exp{ \left( \dfrac{\Delta V_{p,1}}{g_0 I_{\text{sp},l}} \right)}
        \\
        \phi_{a,j} = (\phi^{(d)} \phi^{(l)})_a &= \exp{ \left( \dfrac{\Delta V_{a,2}}{g_0 I_{\text{sp},d}} \right)} 
        \exp{ \left( \dfrac{\Delta V_{a,1}}{g_0 I_{\text{sp},l}} \right)}
    \end{aligned}
\end{equation}
Finally, note that since both $\phi^{(d)}$ and $\phi^{(l)}$ are ratios that do not depend on any mass; hence, $\phi_j$ can be pre-computed once for a given orbital slot and reused regardless of the facility's mass. 
By multiplying the depot's mass at the start of its operation after the arrival insertion burn with $\phi_j$, the EMLEO of the depot is obtained.

\section{Methods}
\label{sec:Methods}
\begin{figure}
    \centering
    \includegraphics[width=0.9\linewidth]{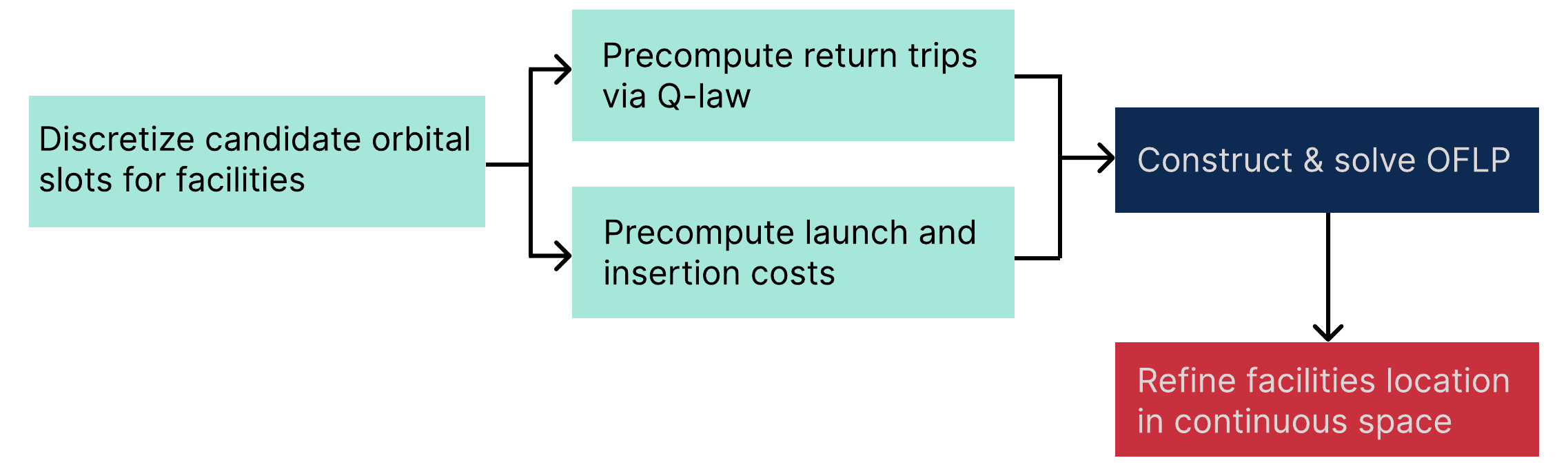}
    \caption{Proposed procedure involving preparing the costs (turquoise), solving the OFLP (navy), and refining in continuous space (red)
    }
    \label{fig:diagram_procedure}
\end{figure}

This section first introduces the static facility location problem (FLP), and discusses its adoption for the orbital facility case, resulting in the orbital facility location problem (OFLP). 
In doing so, discretized candidate locations in space for the facilities must be defined. Considerations involved in this process are discussed in this section. 
Then, the cost coefficients of the FLP must be formulated for the orbital facility case; specifically, the allocation cost consisting of the required propellant for a round-trip between a facility and its client is introduced. 
Finally, once the OFLP is solved, the orbital location of each facility is refined in continuous space, while maintaining the allocations of clients determined by the OFLP. 
Figure \ref{fig:diagram_procedure} illustrates the flow of the proposed method. 

\subsection{Overview of Facility Location Problem}
Given a set of clients $\mathcal{M} = \{1, 2, \ldots, m\}$ and a set of potential sites for facilities $\mathcal{N} = \{1, 2, \ldots, n\}$, the single-source uncapacitated FLP aims to locate $p \leq n$ facilities to service all $m$ clients in a cost-optimal sense. The problem is a binary linear program (BLP), given by
\begin{subequations}
\begin{align}
        \min_{X,Y} \,\, & \,\, \sum_{j=1}^n \boldsymbol{f}_j Y_j + \sum_{i=1}^m \sum_{j=1}^n \boldsymbol{c}_{ij} X_{ij}
        \label{eq:FLP_prob_objective}
\\
        \textit{s.t.} 
        \quad \, & \sum_{j=1}^n X_{ij} = 1 \quad \forall i \in \mathcal{M}
        \label{eq:FLP_prob_constraint_one_allocation}
\\
        &  X_{ij} \leq Y_j \quad \forall i \in \mathcal{M},\, \forall j \in \mathcal{N}
        \label{eq:FLP_prob_constraint_capacity}
\\
        & X_{ij} , \, Y_{j} \in \{0,1\} \quad \forall i \in \mathcal{M} ,\, j \in \mathcal{N}
        \label{eq:FLP_prob_constraint_binary}
\end{align}
\end{subequations}
The decision variables $X \in \mathbb{B}^{m \times n}$ and $Y \in \mathbb{B}^n$ are restricted to be binary by \eqref{eq:FLP_prob_constraint_binary}. The variable $Y_j$ dictates whether the facility $j$ is used within the architecture, and the variable $X_{ij}$ dictates whether facility $j$ is assigned to client $i$. 
The objective of this problem, as given in \eqref{eq:FLP_prob_objective}, is to minimize the sum of costs associated with the allocation of the facility $j$ to client $i$, and the costs of using facility $j$. 
The entries in the matrix $\boldsymbol{c}_{ij}$ are weights associated with the $i$-$j$ allocation, while the entries of the vector $\boldsymbol{f}_j$ are weights associated with using facility $j$. 
Constraints \eqref{eq:FLP_prob_constraint_one_allocation} ensure that exactly one facility is assigned to each client. 
Finally, constraints \eqref{eq:FLP_prob_constraint_capacity} ensure that client $i$ is allocated to a facility $j$ that exists. 
%
A particular advantage of this formulation of the FLP, compared to its continuous analogs, is that the number of facilities does not need to be known a priori. 
Rather, the number of facilities is simply given by $\sum_{j=1}^m Y_j$.

\subsection{Orbital Facility Location Problem}
The FLP is adapted to the case of on-orbit servicing for satellite constellations, called the Orbital Facility Location Problem (OFLP). The OFLP is formulated as follows:
\begin{subequations}
\begin{align}
    \min_{X,Y} \,\, & \,\, 
     \sum_{j=1}^n 
        \underbrace{\mddry \phi_j}_{\boldsymbol{f}_j} Y_j
     +
     \sum_{i=1}^m \sum_{j=1}^n 
        \underbrace{D_i (\tilde{\boldsymbol{c}}_{ij} + \msL) \phi_j}_{\boldsymbol{c}_{ij}} X_{ij} 
    \label{eq:oflp_prob_objective}
    \\
    \textit{s.t.}
    \quad \,
    &  \mddry \phi^{(d)}_{j} Y_j + \sum_{i=1}^m D_i (\tilde{\boldsymbol{c}}_{ij} + \msL) \phi^{(d)}_{j} X_{ij}
    \leq \mlmax 
    \quad \forall j \in \mathcal{N}
    \label{eq:oflp_prob_constraint_launchmass}
    \\&
    \eqref{eq:FLP_prob_constraint_one_allocation}
    -
    \eqref{eq:FLP_prob_constraint_binary}
    \nonumber 
\end{align}
\label{eq:oflp}
\end{subequations}
Here, $\mddry$ is the dry mass of the depot, $\msL$ is the payload mass to be delivered upon each trip to the client satellites, and $D_i$ is the number of trips to be made to each client over the operation period of the depot. 
$\phi_j$ is the mass-ratio of launching the depot to the facility site $j$, computed by equation \eqref{eq:massratio_j_launch_insertion}, and the entries of $\tilde{\boldsymbol{c}}_{ij}$ are the propellant mass required to conduct a return trip from facility $j$ to deliver the payload to client $i$. 
The resulting objective \eqref{eq:oflp_prob_objective} corresponds to the sum of the EMLEO of each facility. 

The OFLP takes the same constraints \eqref{eq:FLP_prob_constraint_one_allocation} - \eqref{eq:FLP_prob_constraint_binary} as the original FLP, with an additional constraint \eqref{eq:oflp_prob_constraint_launchmass} that ensures the wet mass of a given depot does not exceed the maximum launch mass of the launch vehicle $\mlmax$. 
Note that the wet mass of the depot is computed by multiplying only the mass ratio of the depot's burn, $\phi^{(d)}_j$, while the EMLEO in the objective is obtained by multiplying both mass ratios $\phi_j = \phi^{(d)}_j \phi^{(l)}_j$. 

In effect, the OFLP formulation optimizes both the number of facilities as well as their spatial configuration simultaneously through a cost-metric that is standardized in terms of mass. 
The choice of the spatial configuration of a facility results from the trade-off between the cost of establishing it into a particular orbit and the total costs of accessing the allocated clients from it. 

\begin{figure}
    \centering
    \includegraphics[width=0.8\linewidth]{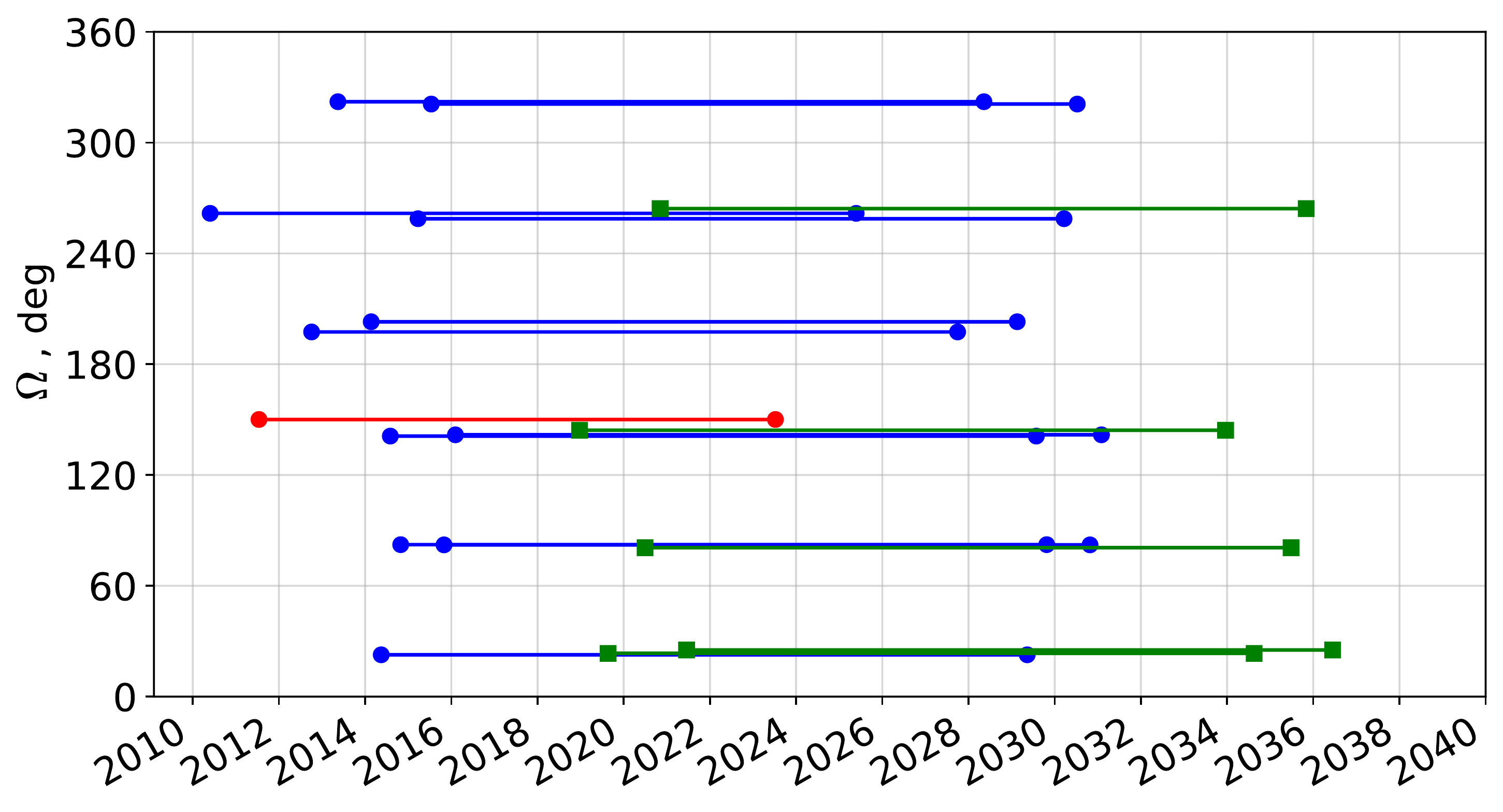}
    \caption{
    Approximate lifetime of currently active GPS Block IIF (blue/red lines, circle markers) and III (green lines, square markers), assuming a lifetime of 15 years from launch, except for Block IIF-2 (red line), which experienced a clock anomaly on 10 July 2023. RAAN corresponds to TLE values queried in December, 2022.
    }
    \label{fig:projected_gps_lifetime}
\end{figure}

Note that the formulation in \eqref{eq:oflp} assumes that the servicer only conducts dedicated servicing round-trips between the depot and a single client satellite, and excludes servicing round-trips that visit multiple client satellites that are ``close'' in orbital elements space, where ``closeness'' is dictated primarily by the orientation of the orbital planes. 
This choice is made to reflect on-orbit servicing needs that occur infrequently within the constellation but with high priority, making the feasibility of servicing multiple client satellites within the same time frame unlikely. 
Typical MEO constellations such as GPS and Galileo fall into this category, as these are expensive assets with life times extending to 15 years. 
For the most recent GPS Block IIF and III, new constituent satellites have been launched at intervals spanning multiple months to a few years within each orbital plane, as shown in Figure \ref{fig:projected_gps_lifetime}. 
Especially for the refueling servicing case, it is thus unlikely that servicing needs would arise at similar time-frames, for example, within a time frame of a few weeks, for two or more client satellites that share a similar orbital plane. 
To avoid having a scenario where a client satellite must wait to be serviced for other nearby client satellite(s) to also require servicing, the depots should be optimally placed considering only dedicated servicing round-trips to one client at a time. 

To consider the off-nominal case where multiple demands on satellites sharing similar orbital planes occur at similar times, we conduct an a posteriori analysis on the cost effectiveness of ``bundling'' multiple servicing needs to a single round-trip; this process is further discussed in Section \ref{sec:multiclient_servicing_trip}. 
Finally, while beyond the scope of this work, a constellation with faster technology refresh rate, typically in LEO, may benefit from servicing round-trips that bundles multiple clients, and would pose a variant of the problem studied in this work.

\subsection{Discretization of Facility Location Candidates}
In the FLP, candidate facility locations must be represented as a discrete number of options in set $\mathcal{N}$. As such, when formulating the OFLP, it is necessary to discretize locations along an orbit into \textit{slots}. To avoid combinatorial explosion, it is necessary to reduce the dimensions of the discretized space. 
Firstly, the anomaly-like variable of the facility is omitted; as discussed earlier, this is a common assumption for multi-revolution low-thrust transfers (see more details in the Q-law literature  \cite{Petropoulos2003,Petropoulos2004,Petropoulos2005}). In our case, the transfers between facilities and clients involve changing any combination of the energy, orbital shape, and orbital plane, within which phasing maneuvers are comparatively inexpensive, especially if the phasing maneuvers can take place over a long time horizon (e.g., when refueling demands for a client satellite can be predicted months in advance). 
As such, the transfer cost between the facilities and the clients may be modeled as an orbital transfer rather than a rendez-vous problem. 
Second, if the constellation shares the same inclination among all of its fleet members, the facility's inclination may be set to match this value. 
Finally, Earth-based constellations are typically near-circular. Then, under the assumption of pure two-body dynamics, the transfer cost from a facility orbit to a client's orbit is negligibly affected by the argument of perigee of the initial departure orbit. 

Hence, making appropriate assumptions that depend on the distribution of the clients, the dimension of the discretization would vary between three and five, among the semimajor axis, eccentricity, inclination, RAAN, and argument of perigee. 
For example in Zhu et al \cite{Zhu2020}, the client satellites are distributed on the same inclination and are near-circular; thus, only the semimajor axis, eccentricity, and RAAN have been considered. 
In the numerical example studied in this work that will be introduced in Section \ref{sec:numerical_results}, the client satellites are near-circular but on varying inclinations; therefore, the facility locations are discretized in terms of semimajor axis, eccentricity, inclination, and RAAN.

\subsection{Computation of Allocation Cost Matrix}
\label{sec:computation_of_allocation_cost_matrix}
\begin{figure}[t]
    \centering
    \includegraphics[width=0.95\linewidth]{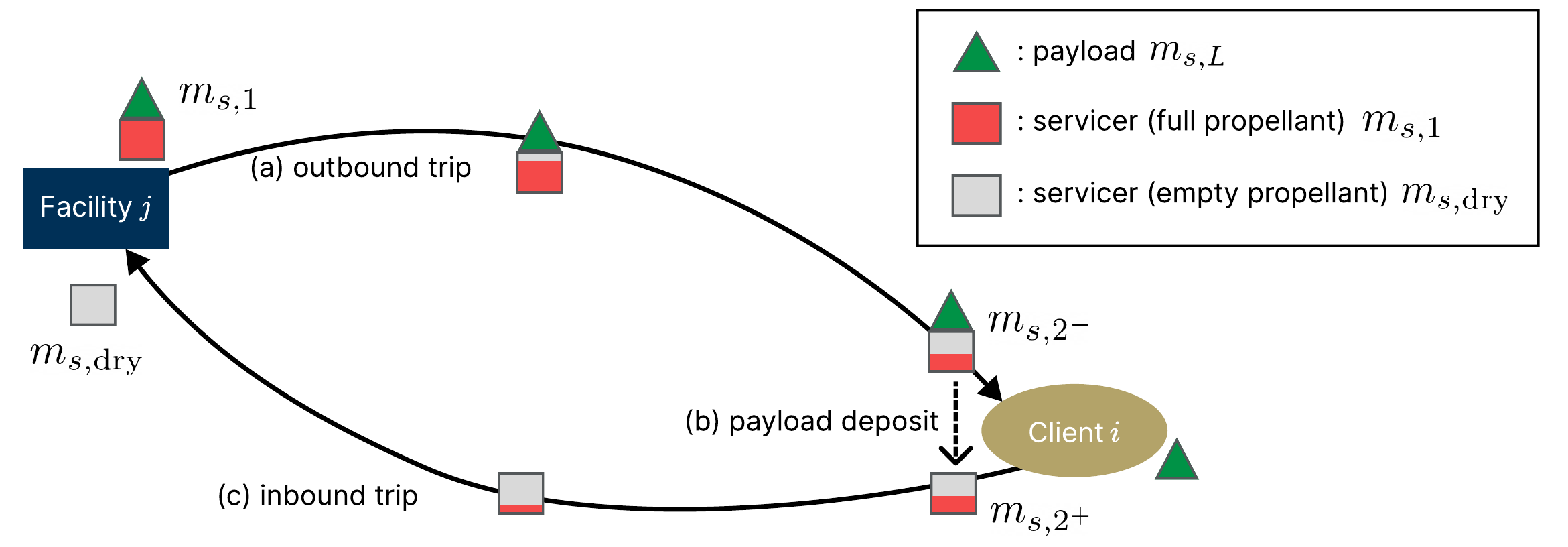}
    \caption{Return trip conducted by servicer between facility $j$ and client $i$. Once the servicer arrives to the client $i$, the payload $\msL$ is deposited.}
    \label{fig:diagram_return_trip}
\end{figure}

The allocation cost matrix $\tilde{\boldsymbol{c}}_{ij}$ has as entries the propellant mass required to allocate client $i$ to facility $j$. This mass is computed by considering the return trip consisting of (a) the outbound trip traveling from the facility to the client satellite, (b) depositing the payload $\msL$ to the client, and (c) the inbound trip returning from the client satellite back to the facility, as illustrated in Figure \ref{fig:diagram_return_trip}. 
Since the overall mass of the spacecraft affects the attainable acceleration and consequently the trajectory as well, the propellant mass expenditures are computed backward in time. 
Note that the mass deposit in step (b) models the unspecified servicing activity, such as a propellant deposit for refueling, or retrofitting an extension module to the client satellite.
The allocation cost is given by
\begin{equation}
    \tilde{\boldsymbol{c}}_{ij} = m_{s,1} - \msdry - \msL
\end{equation}
where $m_{s,1}$ is the servicer's wet mass before departing from the depot and $\msdry$ is the servicer's dry mass.
To obtain the wet mass, the inbound transfer (c) is first solved backward in time, to obtain
\begin{equation}
     \msdry = m_{s,2^+}  - \Delta m_c
\end{equation}
where $m_{s,2^+}$ is the mass of the servicer when it leaves the client, and $\Delta m_c$ is the propellant expenditure on the inbound transfer. 
Then, the outbound transfer (a) can be solved backward in time, such that
\begin{equation}
    m_{s,1} = m_{s,2^-} + \Delta m_a = m_{s,2^+} + \msL + \Delta m_a
\end{equation}
where $m_{s,2^-} = m_{s,2^+} + \msL$ is the mass of the servicer when it arrives at the client, and $\Delta m_a$ is the propellant expenditure on the outbound transfer.
The propellant expenditures in (a) and (c) are computed through Q-law
\begin{equation}
\begin{aligned}
    & \Delta m_a = \dot{m} \Delta t_{\text{OT},ji}
    \\
    & \Delta m_c = \dot{m} \Delta t_{\text{OT},ij}
\end{aligned}
\end{equation}
where $\Delta t_{\text{OT},{ji}}$ is the time taken by the Q-law controller to transfer from facility $j$ to client $i$, $\Delta t_{\text{OT},{ij}}$ is the time taken by the Q-law controller to transfer from client $i$ to facility $j$. 
Note that $\Delta t_{\text{OT},ji} \neq \Delta t_{\text{OT},ij}$ since the mass of the spacecraft during phase (a) and (c) are different. 
Using $\Delta m_a$ and $\Delta m_c$, we may also express $\Tilde{\boldsymbol{c}}_{\boldsymbol{i} j} = \Delta m_a + \Delta m_c$.

It is also possible that the return trip from a particular facility slot to a client is infeasible due to a time of flight that is prohibitively long. In such case, the entry $\tilde{\boldsymbol{c}}_{ij}$ can simply be set to a high value. Alternatively, it is also possible to eliminate the variable $X_{ij}$ by freezing its value to $0$. 
This has the added advantage of reducing the model size and is therefore done in this work.

\subsection{Refinement of Facility Location in Continuous Space}
In OFLP, the orbital slots had to be discretized to efficiently optimize the number of facilities and the allocation of clients to each of them. Thus, given the number of facilities and the allocation of clients obtained from the OFLP solution, we refine the facility locations further in continuous space by formulating a continuous nonlinear programming problem (NLP). 
Unlike OFLP, the refinement problem is solved for each facility separately. Specifically, by collecting the variables dictating the location of a facility into a vector $\aleph$, the refinement problem aims at adjusting the facility's location to minimize its EMLEO, and is given by
\begin{subequations}
    \begin{align}
        \min_{\aleph} \,\, & \,\,\mddry Z(\aleph) 
            + \sum_{i \in \mathcal{M}_a} D_i \left[ \tilde{c}_i(\aleph) + \msL \right] \phi(\aleph)
        \label{eq:refinement_nlp_objective}
        \\
        \textit{s.t.}
        \quad \,
        &  \mddry \phi^{(d)} (\aleph) + \sum_{i \in \mathcal{M}_a} D_i \left[ \tilde{c}_i(\aleph) + \msL \right] \phi^{(d)} (\aleph)
        \leq \mlmax 
        \label{eq:refinement_nlp_constraint}
    \end{align}
    \label{eq:refinement_nlp}
\end{subequations}
where $\mathcal{M}_a$ is the set of indices of clients that are allocated to the given facility. Note also that now, the transfer costs from the facility to each client $\tilde{c}_i$ as well as the facility's mass ratios $\phi$ and $\phi^{(d)}$ are all functions of variables $\aleph$, and must therefore be computed online. 
This online computation of $\tilde{c}_i$ involves calling Q-law during the optimization, which may result in numerical difficulties if gradient-based algorithms are applied to solve \eqref{eq:refinement_nlp}. To circumvent this issue, gradient-free metaheuristic algorithms are employed in this work.

\subsection{Multiclient Servicing Trip}
\label{sec:multiclient_servicing_trip}
As discussed in Section \ref{sec:computation_of_allocation_cost_matrix}, servicing round-trips that visit more than a single client is not included as part of the OFLP \eqref{eq:oflp}. Nevertheless, the cost for such a servicing trip, coined as a ``multiclient servicing trip'', is evaluated to study the feasibility should such an off-nominal need arise. For this portion of the analysis, the facilities' locations are assumed to be fixed from solving the OFLP \eqref{eq:oflp} and the refinement problem \eqref{eq:refinement_nlp}. 

When considering a multiclient servicing trip, the servicing sequence must be determined. This is akin to the traveling salesman problem (TSP), where the initial and final node of the travel loop is the facility. In the context of low-thrust propulsion, the problem is further complicated by the fact that the required propellant mass, corresponding to the edge costs of the TSP, is path-dependent. 
Meanwhile, since MEO constellations have a relatively small number of satellites, and multiclient servicing need is unlikely, the TSP instance consists only of a small number of nodes, so may be solved directly by exhausting all permutations of visits. 

Similarly to the single client-servicing trip, the propellant mass required for a multliclient servicing trip is computed backward in time. 
For a given servicing sequence of $q$ satellites given by $\boldsymbol{i} = [i_1, i_2, \ldots, i_q]$, the arc from the facility $j$ to client $i_q$ is solved backward in time using Q-law, starting with the spacecraft dry mass. This is followed by a payload deposit of mass $\msL$ to client $i_q$ (corresponding to a mass \textit{gain} of the servicer backward in time). Thus far, the computation follows exactly steps (c) and (b), respectively, in Figure \ref{fig:diagram_return_trip}. 
Then, this step is repeated by replacing the arc to go from $i_q$ to $i_{q-1}$, $i_{q-1}$ to $i_{q-2}$, and so on, until the first client $i_1$ is reached. At the end of each arc, the servicer gains the payload mass $\msL$. 
Finally, the outbound trip between the facility $j$ and the first client $i_1$ is computed backward in time. 
The resulting cost for the trip is given by
\begin{equation}
    \Tilde{\boldsymbol{c}}_{\boldsymbol{i} j}
    = \dot{m} 
    \left( 
        \Delta t_{\text{OT},i_q j} 
    + \Delta t_{\text{OT},i_{q-1} i_q}
    + \ldots
    + \Delta t_{\text{OT},i_1 i_2} 
    + \Delta t_{\text{OT}, j i_1} 
    \right)
    \label{eq:multiclient_cost}
\end{equation}
For a given set of clients $\mathcal{I} = \{i_1, i_2,\ldots, i_q\}$, the solution of the TSP is the sequence that minimizes the cost \eqref{eq:multiclient_cost}, denoted as $\boldsymbol{i}^*$, and the corresponding cost is given by
\begin{equation}
    \Tilde{\boldsymbol{c}}_{\boldsymbol{i}^* j}
    = 
    \min_{\boldsymbol{i} \in \boldsymbol{I}} \Tilde{\boldsymbol{c}}_{\boldsymbol{i} j}
    \label{eq:multiclient_cost_optimal}
\end{equation}
where $\boldsymbol{I}$ is the set of permutations of $\mathcal{I}$, consisting of $q!$ elements.

\section{Numerical Results}
\label{sec:numerical_results}
The proposed formulation is applied to design a depot fleet servicing (i) the GPS constellation, (ii) the Galileo constellation, and (iii) these two constellations altogether.
Both the GPS and Galileo constellations are located on Medium Earth Orbits (MEO), at similar ranges of inclination around $53^{\circ}$ to $57^{\circ}$, and a radius of about \SI{26560}{km} and \SI{29600}{km}, respectively. 
The exact orbital elements of the two fleets are given in the Appendix, in Tables \ref{tab:gps_clients} and \ref{tab:galileo_clients}. 
The GPS constellation consists of 31 satellites on six orbital planes, and the Galileo constellation \cite{Bartolomé2015} consists of 28 satellites on four orbital planes. 
Orbital elements of the GPS and Galileo fleets used in this work are given in the Appendix. Figure \ref{fig:clients_distribution} shows the distribution of the clients in inclination and RAAN. 

\begin{figure}
    \centering
    \includegraphics[width=0.9\linewidth]{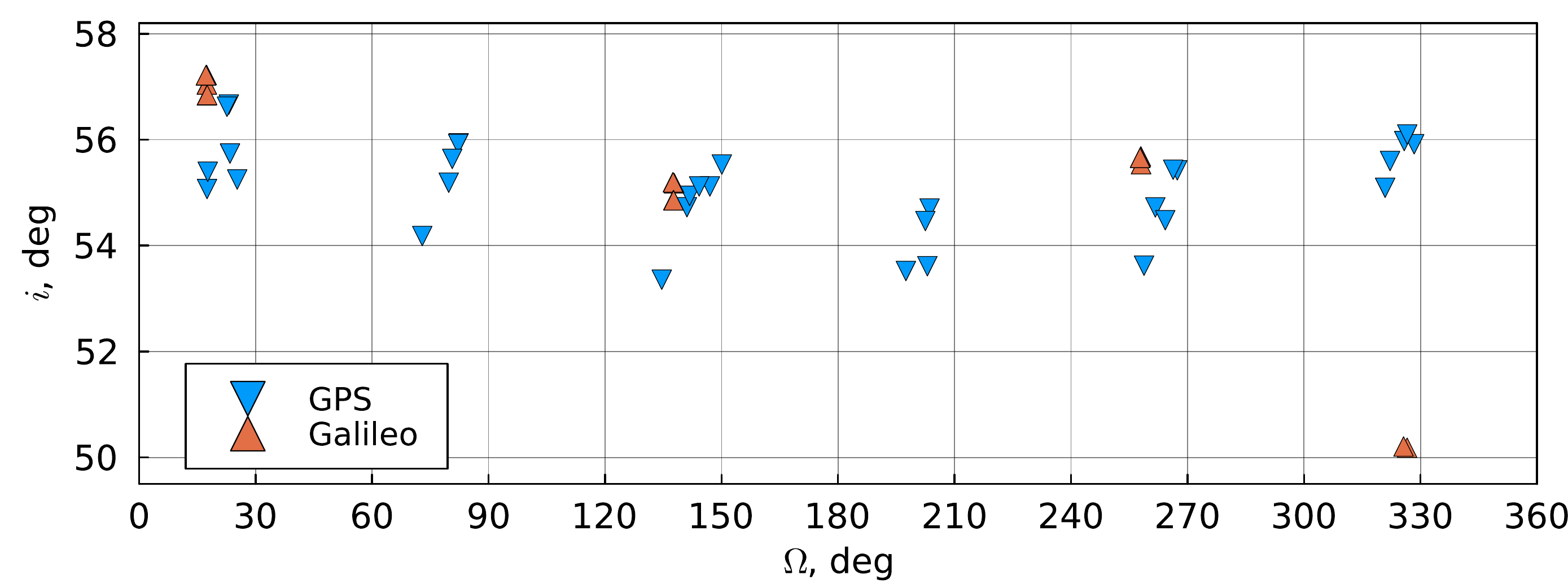}
    \caption{Distribution of GPS and Galileo satellites in inclination and RAAN}
    \label{fig:clients_distribution}
\end{figure}

The OFLP is implemented in Julia and solved with CPLEX 22.1 \cite{cplex2009v12}. 
Table \ref{tab:problem_params} lists the problem parameters used in the numerical experiments. 
The launch vehicle's maximum payload mass capability of $12,950\,\mathrm{kg}$ corresponds to the sub-GEO transfer orbit capability of the Ariane 64 \cite{Ariane6}, used here as an example launch vehicle constraint. 
The servicer vehicle's thrust and $I_{\mathrm{sp},s}$ are taken from Sarton du Jonchay et al \cite{Sarton2021a}. 
Parameters for Q-Law are based on the work by Petropoulos \cite{Petropoulos2005}. 

The numerical results are organized as follows: initially, solutions from OFLP instances of the three scenarios (i) through (iii) are analyzed from a high-level perspective. 
In particular, we provide insight into combining the servicing duties of the two constellations compared to having two individual fleets of depots, one for each constellation. 
Then, a more detailed discussion is provided based on the solutions to scenario (iii), which constitutes the most complex case due to the larger number of client satellites. 
This is followed by a demonstration of the depot location refinement problem \eqref{eq:refinement_nlp_constraint}, which is again applied to scenario (iii). 
Finally, an a posteriori analysis on multiclient trips is provided. 

\begin{table}[h]
\centering
\caption{OFLP and Q-Law parameters for Numerical Experiments}
\begin{tabular}{@{}ll@{}}
\toprule
Parameter                                        & Value(s)            \\ \midrule
Maximum launch mass $\mlmax$, \SI{}{kg}      & 12,950            \\
Launch vehicle parking orbit radius $r_{0}$, \SI{}{km}  & 6,578      \\
Launch vehicle $I_{\mathrm{sp},l}$, \SI{}{sec}   & 457              \\
Payload mass $\msL$, \SI{}{kg}                & 100              \\
Number of trips $D$                              & 1 / 2             \\
Servicer dry mass $\msdry$, \SI{}{kg}  & 500 / 1,000        \\
Servicer $I_{\mathrm{sp},s}$, \SI{}{sec}         & 1,790             \\
Servicer thrust, \SI{}{N}                        & 1.74             \\
Depot dry mass $\mddry$, \SI{}{kg}     & 1,500 / 2,000 / 2,500 \\
Depot $I_{\mathrm{sp},d}$, \SI{}{sec}            & 320              \\ 
Maximum transfer time, \SI{}{days}               & 300              \\
Minimum transfer radius $r_{p \min}$, \SI{}{km}  & 6,878             \\
$W_p$                                            & 1                \\
$W_{\elements}$                                  & [1,1,1,1,1]      \\
$\sigma$                                         & 3                \\
$\nu$                                            & 4                \\
$\zeta$                                          & 2                \\
$k_{r_p}$                                        & 1                \\
\bottomrule
\end{tabular}
\label{tab:problem_params}
\end{table}

\subsection{Depots Placements via Orbital Facility Location Problem}
The candidate orbital slots for the depots are discretized in the four slow Keplerian elements excluding the argument of perigee, $[a,e,i,\Omega]$, as summarized in Table \ref{tab:orbital_slots_gpsgalileo}, resulting in 23,868 slots. 
As previously mentioned, the $\omega$ is not discretized because both the GPS and Galileo satellites are at very low eccentricities. 
While a constant $\omega=0\,\mathrm{deg}$ is assumed for all orbital slots, the resulting solution would be unaffected for any other choice of $\omega$, assuming the transfer cost does not change either. 
In total, the problems has dimensions $X \in \mathbb{B}^{31 \times 23,868}$ and $Y\in \mathbb{B}^{23,868}$, resulting in 763,776 variables when considering the GPS constellation,
$X \in \mathbb{B}^{28 \times 23,868}$ and $Y\in \mathbb{B}^{23,868}$, resulting in 692,172 variables when considering the Galileo constellation, and
$X \in \mathbb{B}^{59 \times 23,868}$ and $Y\in \mathbb{B}^{23,868}$, resulting in 1,432,080 variables when considering both constellations. 

\begin{figure}
    \centering
    \includegraphics[width=0.85\linewidth]{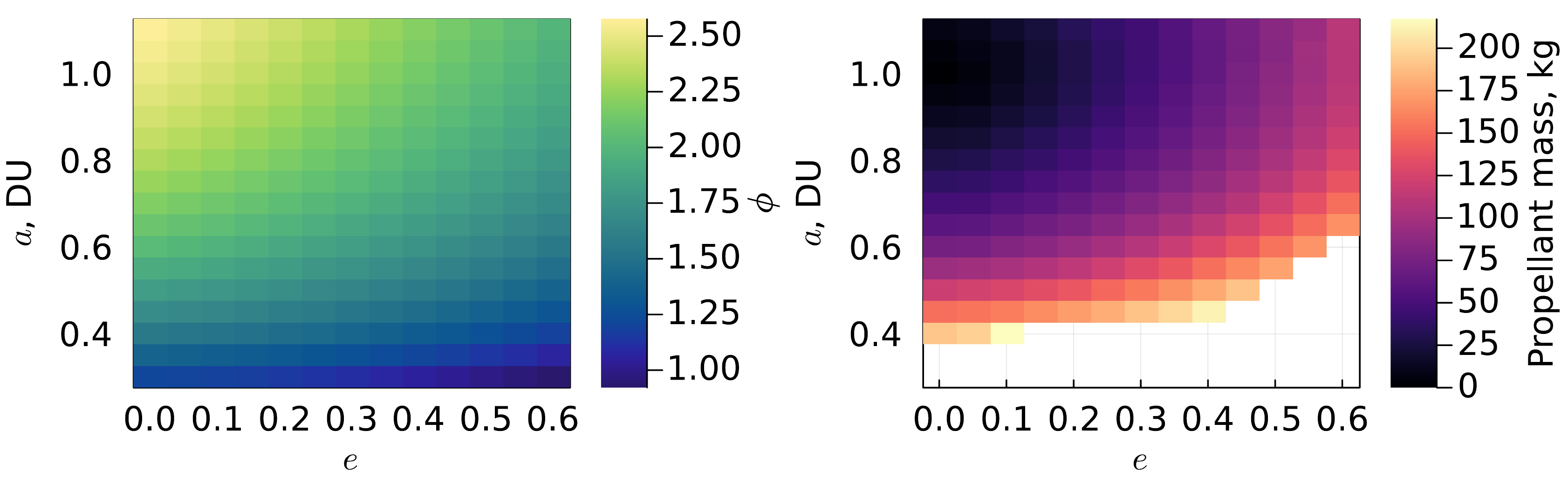}
    \caption{Contour of mass ratio $\phi$ for candidate orbital slots (left) and of transfer propellant mass $\tilde{c}_{ij}$ corresponding to round trip between candidate orbital slots and hypothetical client satellite on circular orbit with $a = 1.0 \, \mathrm{DU}$ and identical orbital plane ($i$ and $\Omega$) as the depot (right)}
    \label{fig:contour_gnss}
\end{figure}

Figure \ref{fig:contour_gnss} shows the contour of facility mass-ratio to $Z$ for the orbital slots in the semimajor axis-eccentricity space, and the contour of allocation cost $\tilde{c}_{ij}$, corresponding to the servicer propellant mass, from the discretized orbital slots to a hypothetical client lying on the same orbital plane, on a circular orbit with a semimajor axis of $1.0$ \SI{}{DU}. 
The missing entries in $\tilde{c}_{ij}$ are due to locations resulting in infeasible transfers for $r_{p \min} = 6878$ \SI{}{km}.  
As expected, the orbital slots closer to the facility yield a lower servicer propellant mass, but tend to come at a larger $Z$, which translates to a larger launch cost. 
The trade-off between these two factors is taken into account by the OFLP. 

\begin{table}[h]
\centering
\caption{Candidate depot orbital slots}
\begin{tabular}{@{}lll@{}}
\toprule
Orbital element          & Value (min:increment:max)       & Number of slots \\ \midrule
Semimajor axis $a$, $\mathrm{DU}$ ($1 \, \mathrm{DU} = 26,560 \,\mathrm{km}$) & 0.3:0.05:1.1 & 17    \\
Eccentricity $e$             & 0:0.05:0.6   & 13              \\
Inclination $i$, deg         & 50:1:58      & 9               \\
RAAN $\Omega$, deg                & 0:30:330     & 12              \\
Argument of perigee $\omega$, deg & 0            & 1               \\ 
Total                    & -            & 23,868           \\
\bottomrule
\end{tabular}
\label{tab:orbital_slots_gpsgalileo}
\end{table}

\begin{table}[]
\caption{
        CPLEX solve time, total EMLEO from optimal solution, and optimal number of depots from for OFLP instances for depots servicing (i) GPS, (ii) Galileo, and (iii) GPS \& Galileo constellations. All problems are solved to optimality with an optimality gap of $0.0$. The \% change in total EMLEO compares case (iii) against the sum of cases (i) and (ii).
        }
\begin{subtable}[h]{\linewidth}
    \centering
    \caption{$\msdry = 500$ \SI{}{kg}}
    \begin{tabular}{@{}llllllll@{}}
    \hline
    \hline
        &       & \multicolumn{3}{c}{D = 1}  & \multicolumn{3}{c}{D = 2}                     \\ \cmidrule(lr){3-5} \cmidrule(l){6-8} 
        & $\mddry$ & (i) GPS & (ii) Galileo & \begin{tabular}[c]{@{}l@{}}(iii) GPS \&\\Galileo (\% change)\end{tabular} & (i) GPS & (ii) Galileo & \begin{tabular}[c]{@{}l@{}}(iii) GPS \&\\Galileo (\% change)\end{tabular} \\ \midrule
        \multirow{3}{*}{\begin{tabular}[c]{@{}l@{}}Solve\\time, sec\end{tabular}}
            & 1,500 & 113.61 & 45.90 & 655.93 & 58.17 & 42.06 & 174.51\\
            & 2,000 & 75.96 & 52.93 & 155.64 & 71.38 & 44.47 & 119.20\\
            & 2,500 & 82.88 & 43.85 & 156.42 & 85.25 & 39.09 & 133.75\\
    \midrule
        \multirow{3}{*}{\begin{tabular}[c]{@{}l@{}}Total\\EMLEO, kg\end{tabular}}
            & 1,500 & 26,290 & 20,148 & 38,079 (-18.0\%) & 38,410 & 32,465 & 59,833 (-15.6\%)\\
            & 2,000 & 29,584 & 22,704 & 42,294 (-19.1\%) & 43,618 & 35,054 & 65,431 (-16.8\%)\\
            & 2,500 & 32,095 & 25,183 & 45,287 (-20.9\%) & 48,451 & 37,694 & 70,865 (-17.7\%)\\
    \midrule
        \multirow{3}{*}{\begin{tabular}[c]{@{}l@{}}Number\\ of depots\end{tabular}}
            & 1,500 & 5 & 3 & 6 & 6 & 3 & 6\\
            & 2,000 & 3 & 3 & 5 & 6 & 3 & 6\\
            & 2,500 & 3 & 3 & 3 & 6 & 3 & 6\\
    \bottomrule
    \end{tabular}
    \label{tab:cross_scenarios_combined_msdry500}
\end{subtable}
\par\bigskip
\begin{subtable}[h]{\linewidth}
    \centering
    \caption{$\msdry = 1,000$ \SI{}{kg}}
    \begin{tabular}{@{}llllllll@{}}
    \hline
    \hline
        &       & \multicolumn{3}{c}{D = 1}                      & \multicolumn{3}{c}{D = 2}                     \\ \cmidrule(lr){3-5} \cmidrule(l){6-8} 
        & $\mddry$ & (i) GPS & (ii) Galileo & \begin{tabular}[c]{@{}l@{}}(iii) GPS \&\\Galileo (\% change)\end{tabular} & (i) GPS & (ii) Galileo & \begin{tabular}[c]{@{}l@{}}(iii) GPS \&\\Galileo (\% change)\end{tabular} \\ \midrule
        \multirow{3}{*}{\begin{tabular}[c]{@{}l@{}}Solve\\time, sec\end{tabular}}
            & 1,500 & 71.55 & 50.87 & 295.82 & 56.53 & 38.55 & 257.04\\
            & 2,000 & 127.40 & 63.08 & 461.12 & 63.66 & 61.41 & 281.12\\
            & 2,500 & 66.84 & 42.66 & 787.71 & 70.55 & 53.93 & 2709.11\\
    \midrule
        \multirow{3}{*}{\begin{tabular}[c]{@{}l@{}}Total\\EMLEO, kg\end{tabular}}
            & 1,500 & 32,010 & 26,641 & 47,019 (-19.8\%) & 45,817 & 42,632 & 74,732 (-15.5\%)\\
            & 2,000 & 37,564 & 29,336 & 53,192 (-20.5\%) & 52,263 & 46,573 & 81,195 (-17.8\%)\\
            & 2,500 & 42,730 & 31,996 & 58,786 (-21.3\%) & 58,377 & 50,257 & 90,427 (-16.8\%)\\
    \midrule
        \multirow{3}{*}{\begin{tabular}[c]{@{}l@{}}Number\\ of depots\end{tabular}}
            & 1,500 & 6 & 3 & 6 & 6 & 4 & 6\\
            & 2,000 & 6 & 3 & 6 & 6 & 4 & 6\\
            & 2,500 & 6 & 3 & 6 & 6 & 4 & 6\\
    \bottomrule
    \end{tabular}
    \label{tab:cross_scenarios_combined_msdry1000}
\end{subtable}
    \label{tab:cross_scenarios_combined}
\end{table}

The effect of varying the demand $D$, the servicer dry mass $\msdry$, and the depot dry mass $\mddry$ are investigated by constructing and solving OFLP instances for all possible combinations of these parameters and all three scenarios (i) through (iii).
Table \ref{tab:cross_scenarios_combined} summarizes the performance of CPLEX solve time on a computer with 8 core i7-10700 CPU and 32 GB computer memory as well as the total $\mathrm{EMLEO}$ and the number of depots from the optimal solution, against $\msdry$. 
Looking at the solve times, in general, larger problem instances (in increasing order, (ii), (i), followed by (iii)) lead to longer computation time. 
The spike seen in solve time for scenario (iii), $D = 2$, and $\msdry = 1,000$ \SI{}{kg} on the top left window is due to the launch mass constraint \eqref{eq:oflp_prob_constraint_launchmass} that prohibits a solution akin to the optimal solutions in the other OFLP instances.

The trend in the total $\mathrm{EMLEO}$ is monotonic and nearly linear against $\mddry$, as well as a monotonic increase for larger constellations, for larger $\msdry$, or more demand $D$; this is a direct consequence of the objective function \eqref{eq:oflp_prob_objective} being monotonic to the aforementioned parameters. 
Of particular note is the combined total $\mathrm{EMLEO}$ from scenario (iii), which is less than the combined total $\mathrm{EMLEO}$ from scenarios (i) and (iii) by around 15\% to 20\%. 
This is indicative of the fact that the difference in semimajor axes between the two constellations is small enough that launching depots servicing both GPS and Galileo satellites simultaneously leads to savings in $\mathrm{EMLEO}$. 

Finally, for solutions involving GPS (scenarios (i) and (iii)), all cases except for scenario (i) $D=1$, $\mddry = 1500$ \SI{}{kg} and scenario (iii) $D=1$, $\mddry = 2000$ \SI{}{kg} involve 3 or 6 depots; these get allocated to the constellations spanning 6 orbital planes as shown in Figure \ref{fig:clients_distribution}, where clients in either 1 or 2 orbital planes are allocated to each depot. 
The two aforementioned solutions involving 5 depots consist of a combination of 1 depot allocated to clients over 2 orbital planes and 4 depots each allocated to clients on respective orbital planes. 
The solutions from scenario (ii) consistently involve either 3 or 4 depots; the Galileo constellation spans four orbital planes, but two among these planes are relatively close to one another, with a difference in RAAN of about $60^{\circ}$; thus, as long as the wet mass of each depot is not prohibitively large (as is the case for $\msdry = 1000$ \SI{}{kg} and $D = 2$, which results in 4 depots), the optimal solution bundles these two orbital planes to a single depot. 

\begin{landscape}
\begin{table}[!]
\centering
\caption{Summary of optimal GPS \& Galileo servicing depots from OFLP instances with $\msdry = 500$ \SI{}{kg}}
\begin{tabular}{lllllllllllllllllll}
\hline
\hline
  & & \multicolumn{7}{c}{$D = 1$} &  & \multicolumn{7}{c}{$D = 2$}   \\
  \cline{3-10} \cline{12-19}
\begin{tabular}[c]{@{}l@{}} $\mddry$, \\ \SI{}{kg}  \end{tabular}
& No. & 
\begin{tabular}[c]{@{}l@{}} $a$, \\ \SI{}{DU} \end{tabular} &
$e$ & 
\begin{tabular}[c]{@{}l@{}} $i$,\\ \SI{}{deg} \end{tabular}  &
\begin{tabular}[c]{@{}l@{}} $\Omega$, \\ \SI{}{deg} \end{tabular}  &
$\mathcal{M}_a$  & 
\begin{tabular}[c]{@{}l@{}} $\mdwet$, \\ \SI{}{kg}  \end{tabular} &
\begin{tabular}[c]{@{}l@{}} Depot \\ EMLEO, \SI{}{kg}  \end{tabular} &
\begin{tabular}[c]{@{}l@{}} Total \\ EMLEO, \SI{}{kg} \end{tabular} 
&  &   
\begin{tabular}[c]{@{}l@{}} $a$ \\ \SI{}{DU} \end{tabular} &
$e$ &
\begin{tabular}[c]{@{}l@{}} $i$ \\ \SI{}{deg} \end{tabular}  &
\begin{tabular}[c]{@{}l@{}} $\Omega$ \\ \SI{}{deg} \end{tabular}  &
$\mathcal{M}_a$  & 
\begin{tabular}[c]{@{}l@{}} $\mdwet$ \\ \SI{}{kg}  \end{tabular} &
\begin{tabular}[c]{@{}l@{}} Depot \\ EMLEO, \SI{}{kg}  \end{tabular} &
\begin{tabular}[c]{@{}l@{}} Total \\ EMLEO, \SI{}{kg} \end{tabular} 
\\[0.8em]
\hline
\multirow{6}{*}{1,500} 
& 1 & 0.80 & 0.20 &  56 &  30 & 16 &  6,015 &  9,470  & \multirow{6}{*}{38,079} & & 0.95 & 0.10 &  55 &  30 & 16 &  9,612 & 15,416 & \multirow{6}{*}{59,833}\\
& 2 & 0.60 & 0.55 &  57 &  90 &  5 &  2,758 &  4,312 & & & 0.60 & 0.55 &  57 &  90 &   5 & 3,973 &  6,211 \\
& 3 & 0.60 & 0.55 &  55 & 150 & 14 &  4,845 &  7,575 & & & 0.60 & 0.55 &  55 & 150 & 14 &  8,147 & 12,737 \\
& 4 & 0.55 & 0.50 &  54 & 210 &  4 &  2,501 &  3,799 & & & 0.60 & 0.55 &  53 & 210 &  4 &  3,343 &  5,227 \\
& 5 & 0.60 & 0.55 &  57 & 270 & 13 &  4,682 &  7,320 & & & 0.60 & 0.55 &  57 & 270 & 13 &  7,821 & 12,226 \\
& 6 & 0.55 & 0.50 &  54 & 330 &  7 &  3,688 &  5,603 & & & 0.95 & 0.05 &  55 & 330 &  7 &  5,048 &  8,016 \\
\hline
\multirow{6}{*}{2,000} 
& 1 & 0.75 & 0.25 &  55 &  30 & 16 &  6,734 & 10,546 & \multirow{5}{*}{42,294} & & 0.90 & 0.10 &  55 &  30 & 16 & 10,475 & 16,606 & \multirow{6}{*}{65,431} \\
& 2 & 0.40 & 0.35 &  53 &  90 &  5 &  3,726 &  5,022 & & & 0.60 & 0.55 &  57 &  90 &  5 &  4,487 &  7,015 \\
& 3 & 0.60 & 0.55 &  55 & 150 & 14 &  5,360 &  8,379 & & & 0.60 & 0.55 &  55 & 150 & 14 &  8,662 & 13,541 \\
& 4 & 0.65 & 0.60 &  51 & 240 & 17 &  7,473 & 11,973 & & & 0.55 & 0.50 &  54 & 210 &  4 &  3,963 &  6,021 \\
& 5 & 0.50 & 0.45 &  53 & 330 &  7 &  4,338 &  6,374 & & & 0.60 & 0.55 &  57 & 270 & 13 &  8,335 & 13,031 \\
& 6 & \multicolumn{7}{c}{Unused}      & & & 0.90 & 0.10 &  54 & 330 &  7 &  5,814 &  9,216 \\
\hline
\multirow{6}{*}{2,500} 
& 1 & 0.90 & 0.55 &  53 &   0 & 23 & 10,724 & 18,172 & \multirow{3}{*}{45,287} & & 0.90 & 0.10 &  55 &  30 & 16 & 11,215 & 17,779 & \multirow{6}{*}{70,865} \\
& 2 & 0.65 & 0.60 &  52 & 120 & 19 &  8,942 & 14,328 & & & 0.60 & 0.55 &  57 &  90 &  5 &  5,002 &  7,819 \\
& 3 & 0.65 & 0.60 &  51 & 240 & 17 &  7,980 & 12,786 & & & 0.60 & 0.55 &  55 & 150 & 14 &  9,176 & 14,346 \\
& 4 &  \multicolumn{7}{c}{Unused}     & & & 0.55 & 0.50 &  54 & 210 &  4 &  4,482 &  6,810 \\
& 5 &  \multicolumn{7}{c}{Unused}     & & & 0.60 & 0.55 &  57 & 270 & 13 &  8,850 & 13,835 \\
& 6 &  \multicolumn{7}{c}{Unused}     & & & 0.65 & 0.25 &  54 & 330 &  7 &  6,790 & 10,276\\ 
\hline
\hline
\end{tabular}
\label{tab:used_elts_FLP_gpsgal_ms500}
\end{table}

\begin{table}[!]
\centering
\caption{Summary of optimal GPS \& Galileo servicing depots from OFLP instances with $\msdry = 1,000$ \SI{}{kg}}
\begin{tabular}{lllllllllllllllllll}
\hline
\hline
  & & \multicolumn{7}{c}{$D = 1$} &  & \multicolumn{7}{c}{$D = 2$}   \\
  \cline{3-10} \cline{12-19}
\begin{tabular}[c]{@{}l@{}} $\mddry$, \\ \SI{}{kg}  \end{tabular}
& No. & 
\begin{tabular}[c]{@{}l@{}} $a$, \\ \SI{}{DU} \end{tabular} &
$e$ & 
\begin{tabular}[c]{@{}l@{}} $i$,\\ \SI{}{deg} \end{tabular}  &
\begin{tabular}[c]{@{}l@{}} $\Omega$, \\ \SI{}{deg} \end{tabular}  &
$\mathcal{M}_a$  & 
\begin{tabular}[c]{@{}l@{}} $\mdwet$, \\ \SI{}{kg}  \end{tabular} &
\begin{tabular}[c]{@{}l@{}} Depot \\ EMLEO, \SI{}{kg}  \end{tabular} &
\begin{tabular}[c]{@{}l@{}} Total \\ EMLEO, \SI{}{kg} \end{tabular} 
&  &   
\begin{tabular}[c]{@{}l@{}} $a$ \\ \SI{}{DU} \end{tabular} &
$e$ &
\begin{tabular}[c]{@{}l@{}} $i$ \\ \SI{}{deg} \end{tabular}  &
\begin{tabular}[c]{@{}l@{}} $\Omega$ \\ \SI{}{deg} \end{tabular}  &
$\mathcal{M}_a$  & 
\begin{tabular}[c]{@{}l@{}} $\mdwet$ \\ \SI{}{kg}  \end{tabular} &
\begin{tabular}[c]{@{}l@{}} Depot \\ EMLEO, \SI{}{kg}  \end{tabular} &
\begin{tabular}[c]{@{}l@{}} Total \\ EMLEO, \SI{}{kg} \end{tabular} 
\\[0.8em]
\hline
\multirow{6}{*}{1,500} 
& 1 & 1.00 & 0.10 &  55 &  30 & 16 &  7,028 & 11,392 & \multirow{6}{*}{47,019} & & 1.00 & 0.10 &  55 &  30 & 16 & 11,811 & 19,146 & \multirow{6}{*}{74,732} \\
& 2 & 0.60 & 0.55 &  58 &  90 &  5 &  3,377 &  5,279 & & & 0.60 & 0.55 &  58 &  90 &  5 &  5,211 &  8,146 \\
& 3 & 0.95 & 0.10 &  55 & 150 & 14 &  6,208 &  9,956 & & & 0.95 & 0.10 &  55 & 150 & 14 & 10,183 & 16,331 \\
& 4 & 0.60 & 0.55 &  53 & 210 &  4 &  2,886 &  4,513 & & & 0.60 & 0.55 &  53 & 210 &  4 &  4,230 &  6,612 \\
& 5 & 0.95 & 0.15 &  55 & 270 & 13 &  5936 &  9,609 & & & 1.00 & 0.10 &  55 & 270 & 13 &  9,659 & 15,658 \\
& 6 & 0.95 & 0.05 &  55 & 330 &  7 &  3,948 &  6,270 & & & 1.00 & 0.00 &  56 & 330 &  7 &  5,564 &  8,839 \\
\hline
\multirow{6}{*}{2,000} 
& 1 & 0.95 & 0.10 &  55 &  30 & 16 &  7,849 & 12,589 & \multirow{6}{*}{53,192} & & 1.00 & 0.10 &  55 &  30 & 16 & 12,559 & 20,358  & \multirow{6}{*}{81,195}
\\
& 2 & 0.60 & 0.55 &  58 &  90 &  5 &  3,891 &  6,084 & & & 0.60 & 0.55 &  58 &  90 &  5 &  5,725 &  8,950 
\\
& 3 & 0.60 & 0.55 &  55 & 150 & 14 &  7,061 & 11,038 & & & 0.95 & 0.10 &  55 & 150 & 14 & 10,927 & 17,525 
\\
& 4 & 0.55 & 0.50 &  54 & 210 &  4 &  3,498 &  5,315  & & & 0.60 & 0.55 &  53 & 210 &  4 &  4,744 &  7,416 
\\
& 5 & 0.60 & 0.55 &  56 & 270 & 13 &  6,833 & 10,682 & & & 0.95 & 0.10 &  55 & 270 & 13  & 10,507 & 16,852 
\\
& 6 & 0.95 & 0.05 &  55 & 330 &  7 &  4,713 &  7,484  & & & 1.00 & 0.00 &  56 & 330 &  7  &  6,353 & 10,093 
\\
\hline
\multirow{6}{*}{2,500} 
& 1 & 0.95 & 0.10 &  55 &  30 & 16 &  8,594 & 13,783  & \multirow{6}{*}{58,786} & & 1.00 & 0.10 &  55 &  30 & 15 & 12,653 & 20,510  & \multirow{6}{*}{90,427} \\
& 2 & 0.60 & 0.55 &  58 &  90 &  5 &  4,406 &  6,888 & & & 0.60 & 0.55 &  58 &  90 &  5 &  6,240 &  9,754  \\
& 3 & 0.60 & 0.55 &  55 & 150 & 14 &  7,575 & 11,842  & & & 0.95 & 0.10 &  55 & 150 & 14 & 11,671 & 18,719  \\
& 4 & 0.55 & 0.50 &  54 & 210 &  4 &  4,018 &  6,104  & & & 0.60 & 0.55 &  53 & 210 &  4 &  5,258 &  8,221  \\
& 5 & 0.60 & 0.55 &  56 & 270 & 13 &  7,348 & 11,487  & & & 0.95 & 0.10 &  55 & 270 & 13 & 11,252 & 18,046 \\
& 6 & 0.90 & 0.10 &  55 & 330 &  7 &  5,477 &  8,682  & & & 1.00 & 0.05 &  56 & 330 &  8 &  9,453 & 15,176  \\ 
\hline
\hline
\end{tabular}
\label{tab:used_elts_FLP_gpsgal_ms1000}
\end{table}
\end{landscape}

\begin{figure}[h!]
    \centering
    \includegraphics[width=0.9\linewidth]{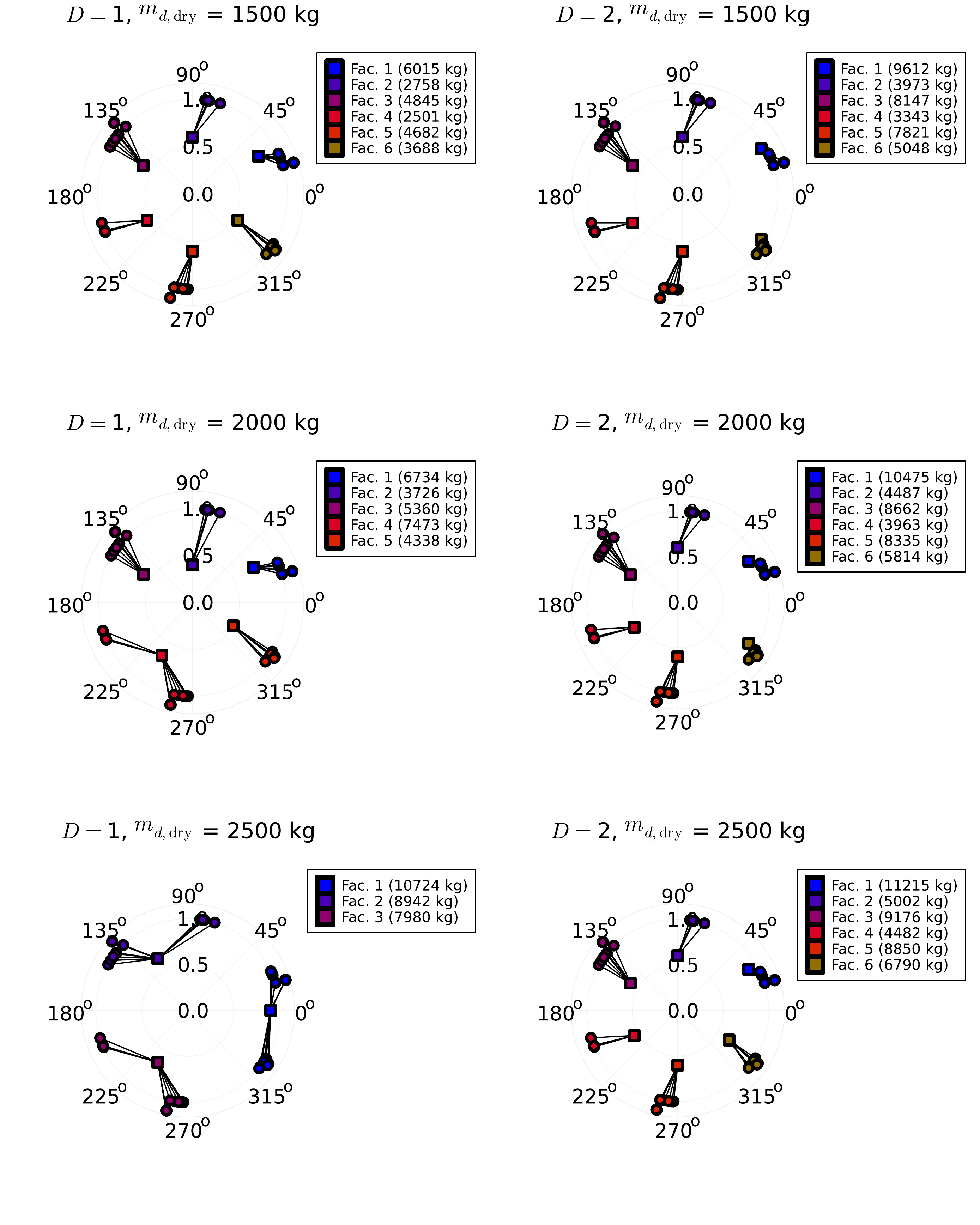}
    \caption{GPS \& Galileo satellite (circles) and facility (squares) locations from OFLP in $(a \,[\mathrm{DU}],\Omega\,[\mathrm{deg}])$ space labeled in terms of depot wet mass, with $\msdry=500\,\mathrm{kg}$}
    \label{fig:pcomb_gnss_euus_prob201_loose}
\end{figure}

\begin{figure}[h!]
    \centering
    \includegraphics[width=0.975\linewidth]{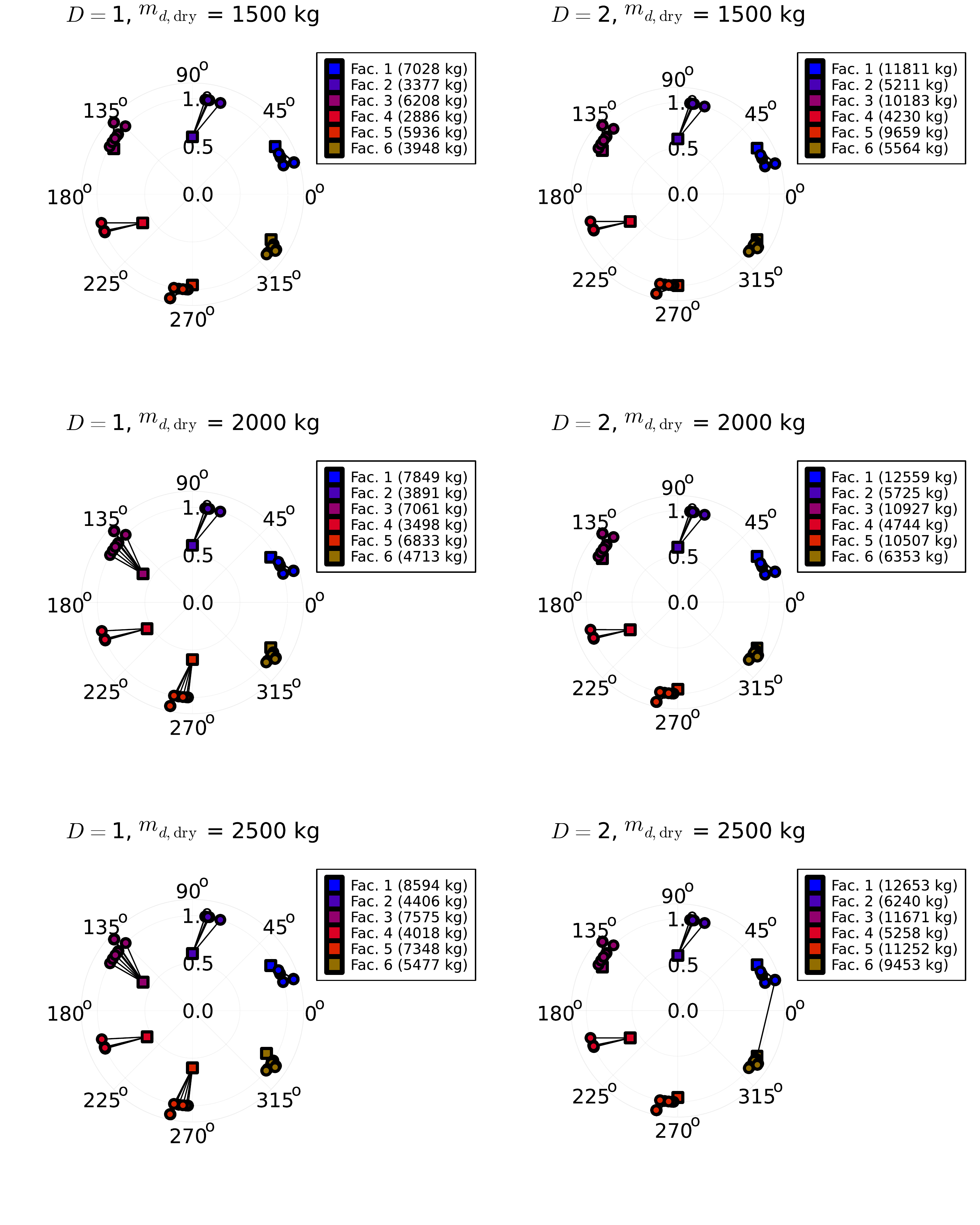}
    \caption{GPS \& Galileo satellite (circles) and facility (squares) locations from OFLP in $(a \,[\mathrm{DU}],\Omega\,[\mathrm{deg}])$ space labeled in terms of depot wet mass, with $\msdry=1,000\,\mathrm{kg}$}
    \label{fig:pcomb_gnss_euus_prob202_loose}
\end{figure}

\begin{figure}[h!]
    \centering
    \includegraphics[width=0.975\linewidth]{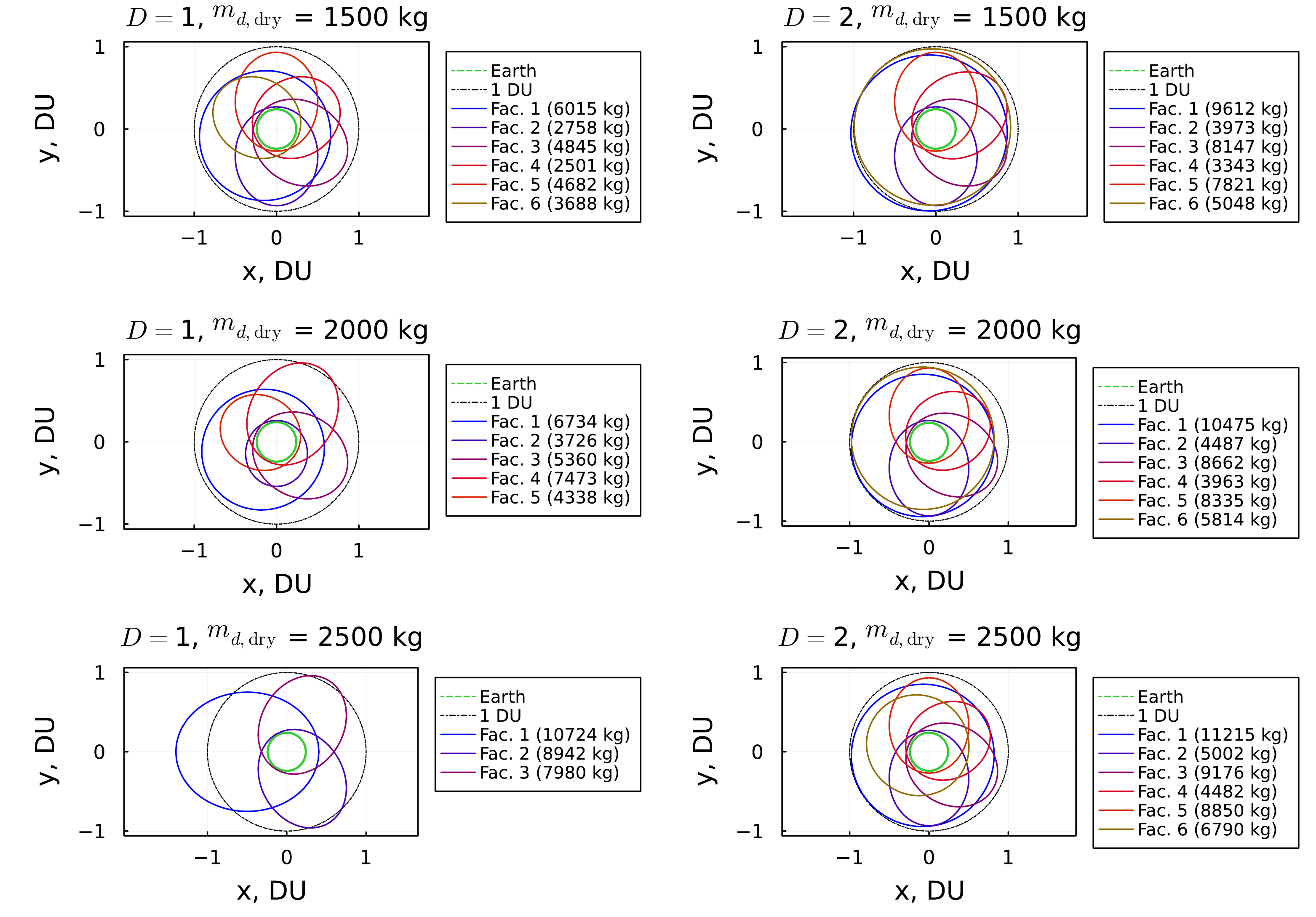}
    \caption{GPS \& Galileo servicing facility orbits from OFLP with $\msdry=500\,\mathrm{kg}$, shown in perifocal frame projection with perigee direction based on $\Omega$, where perigee along positive $x$-axis corresponds to $\Omega = 0$, labeled in terms of depot wet mass}
    \label{fig:ppf_gnss_euus_prob201_loose}
\end{figure}

\begin{figure}[h!]
    \centering
    \includegraphics[width=0.9\linewidth]{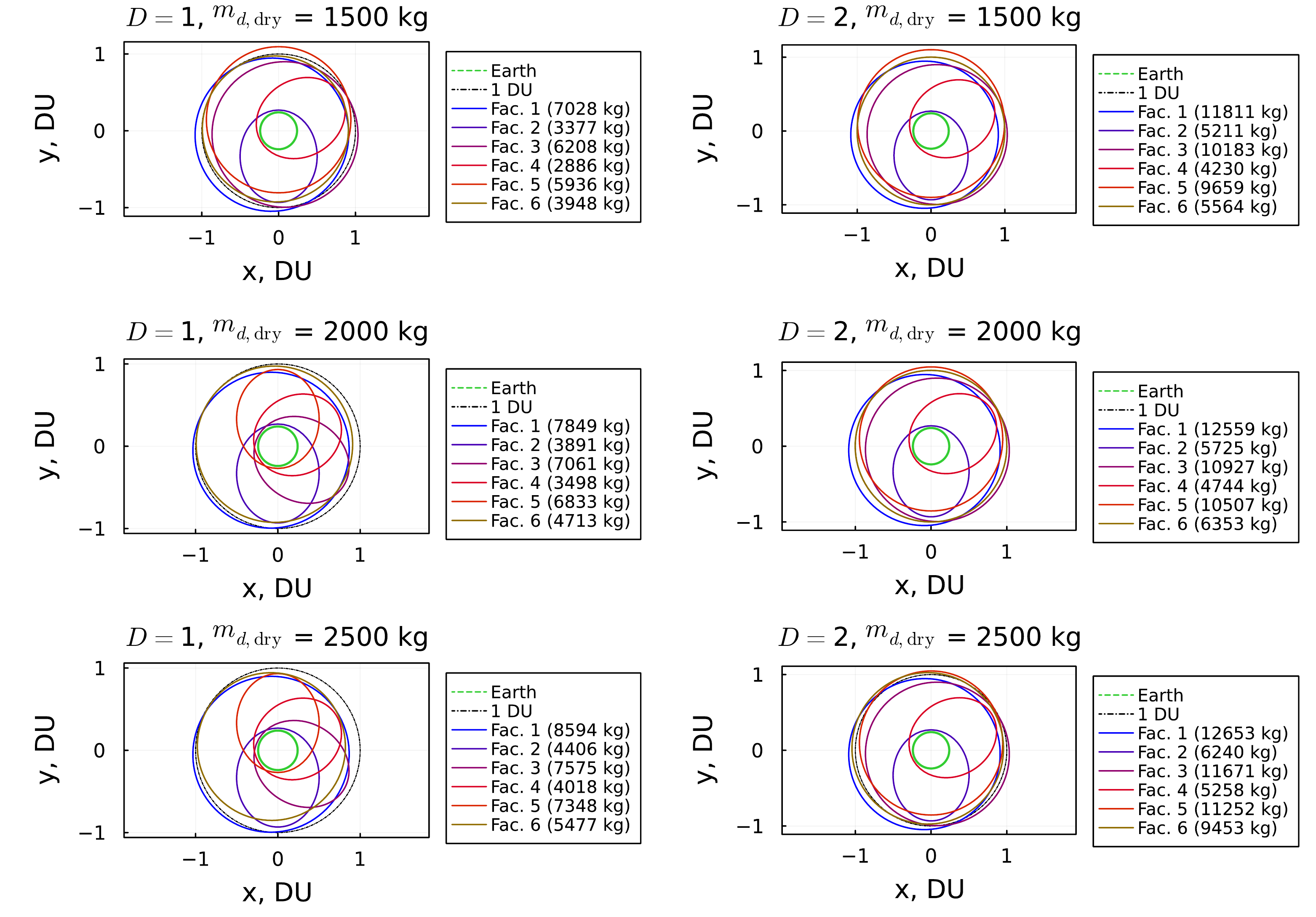}
    \caption{GPS \& Galileo servicing facility orbits from OFLP, shown in perifocal frame projection with perigee direction based on $\Omega$, where perigee along positive $x$-axis corresponds to $\Omega = 0$, with $\msdry=1,000\,\mathrm{kg}$, labeled in terms of depot wet mass}
    \label{fig:ppf_gnss_euus_prob202_loose}
\end{figure}

Due to the relative complexity of the OFLP instance arising from a larger number of client satellites as well as the efficacy of utilizing depots for both constellations simultaneously, the subsequent analysis is based on scenario (iii), for depots catering to both GPS and Galileo fleets.
Figures \ref{fig:pcomb_gnss_euus_prob201_loose} and \ref{fig:pcomb_gnss_euus_prob202_loose} show the distribution of facilities in semimajor axis-RAAN space.
Figures \ref{fig:ppf_gnss_euus_prob201_loose} and \ref{fig:ppf_gnss_euus_prob202_loose} show the orbits of the facilities in each on their perifocal frame rotated about the $z$-axis by $\Omega$. 
The full orbital elements, wet mass, and number of clients allocated to each depot are provided in Tables \ref{tab:used_elts_FLP_gpsgal_ms500} and \ref{tab:used_elts_FLP_gpsgal_ms1000}. 

Some intuitions can be obtained from these results. 
Firstly, as expected, facilities tend to align with the clusters of clients, located at 6 distinct regions in $\Omega$-space, as shown in Figures \ref{fig:pcomb_gnss_euus_prob201_loose} and \ref{fig:pcomb_gnss_euus_prob202_loose}. 
The slight misalignment between the facility and the centroid of the cluster is an artifact of the discretization of orbital slots. 

Also, as it is visible in both Figures \ref{fig:ppf_gnss_euus_prob201_loose} and \ref{fig:ppf_gnss_euus_prob202_loose}, orbits of facilities can be categorized into ellipses with low eccentricities of around $0.2$ to $0.35$ and ellipses with higher eccentricities of $0.5$ or higher. For the same semimajor axis, the former type requires a larger facility location mass ratio $Z$ but the servicer transfer cost is lower, while the latter type has a lower $Z$ and higher servicer transfer cost. 
Notably, facilities servicing multiple locations in RAAN, namely facility 4 in $D=1$ and $\mddry = 2,000$ \SI{}{kg} as well as all three facilities in $D=1$ and $\mddry = 2,500$ \SI{}{kg}, have relatively high eccentricities. 

Finally, Figure \ref{fig:pcomb_gnss_euus_prob202_loose} immediately reveals a noteworthy allocation for $D=2$ and $\mddry = 2,500$ \SI{}{kg} to facility 6 at $\Omega = 330$ \SI{}{deg}, stretching a client in the cluster close to $\Omega = 30$ \SI{}{deg}. 
This is due to the maximum launch mass constraint \eqref{eq:oflp_prob_constraint_launchmass} which restricts the wet mass of the facility from exceeding $\mlmax$. 
Indeed, facility 1 at $\Omega = 30$ \SI{}{deg} has a wet mass of $12,653$ \SI{}{kg}, which is very close to $\mlmax = 12,950$ \SI{}{kg}.  
As a result, facility 6 takes up an additional $8^{\mathrm{th}}$ client, as seen in Table \ref{tab:used_elts_FLP_gpsgal_ms1000}. 
It is also noted that this solution with a divided allocation of clients in a cluster to multiple facilities is more challenging to CPLEX, and requires a longer solve time of approximately 45 minutes, compared to the other solutions which are found in up to around 10 minutes, as indicated in Table \ref{tab:cross_scenarios_combined_msdry1000}.

In the following subsections, the effect of the three parameters, $D$, $\msdry$, and $\mddry$, is analyzed in further detail. 

\subsubsection{Effect of Facility Dry Mass}
An increase in facility dry mass $\mddry$ means an increase in facility launch cost; two things are noted with respect to increasing $\mddry$.
Firstly, looking at facilities located at the same RAAN, an increase in $\mddry$ is seen to result in some cases to a decrease in $a$ and an increase in $e$, provided that the same clients are still allocated to the specific depot. 
This is due to the lower $Z$ that can be achieved when increasing $e$, as visible from the contour of $Z$ in Figure \ref{fig:contour_gnss}. 
While decreasing $a$ would also decrease $Z$, decreasing the energy of the facility leads to a more significant penalty on the transfer cost $\tilde{c}_{ij}$ than the penalty associated with increasing $e$ of the facility. 
This is seen for example in Figure \ref{fig:pcomb_gnss_euus_prob201_loose} for the depot at $\Omega = 30 \,\mathrm{deg}$ for $D=1$, or the depot at $\Omega = 330\,\mathrm{deg}$ with $D=2$. 
However, more commonly, the depot locations are observed to be mostly unaffected by an increase in $\mddry$. 
This variability of behavior can be attributed to the discretization of the facility slots; if the benefit of moving a depot is too small to be captured by adjacent facility slots, the optimal solution would not move the depot. It can thus be discerned that the corresponding trend is relatively weak for the chosen coarseness of the facility slots. 

Secondly, an increase in $\mddry$ is also observed to occasionally alter the distribution of the clients among the depots as well. 
Optimal solutions bundling two clusters of clients into a single facility begin to appear, as long as the facility mass does not exceed the maximum launch mass, and the required number of return trips dictated by $D$ does not increase. 
This is observed for $D=1$, where the clusters of clients in the fourth quadrant of $\Omega$ are bundled when increasing $\mddry$ from \SI{1,500}{kg} to \SI{2,000}{kg}, and the other four clusters are also bundled when increasing $\mddry$ further to \SI{2,500}{kg}.

It is also noteworthy that in these cases, a facility that services multiple clusters of clients is consistently located at an inclination of $53$ \SI{}{deg} or lower, despite the fact that all clients except for one have inclinations of $53$ \SI{}{deg} or higher. 
This is a result of the transfer cost involved when the servicer has to make significant changes in RAAN during the round-trip; a lower inclination is more beneficial for such a maneuver providing a larger ``lever-arm'' to rotate the orbital plane, even though there will be added cost for having to adjust the inclination to match the clients'.

\subsubsection{Effect of Demand}
The demand $D$ reflects the number of round trips to be conducted between a depot and its clients. While this does not affect the cost of the transfer itself, the additional round trip requires additional propellant and payload to be brought and stored at the depot. As such, as $D$ increases, the saving in propellant mass resulting from placing the depot closer to the client increases. 
For the depots at the same $\Omega$ between $D=1$ and $D=2$, the orbits of the facilities are generally moved ``closer''  to the clients, by either increasing $a$, reducing $e$, or both. 
Examples of such depots include the one at $\Omega = 30$ \SI{}{deg} in all cases except for $\mddry = 1,500$ \SI{}{kg} and $\msdry = 1,000$ \SI{}{kg}, as presented in Tables \ref{tab:used_elts_FLP_gpsgal_ms500} and \ref{tab:used_elts_FLP_gpsgal_ms1000}. 
The existence of some depots remaining in the slots despite the increase in $D$ can again be explained by the coarseness of the discretization of the facility slots.

\subsubsection{Effect of Servicer Dry Mass}
Increasing the servicer dry mass $\msdry$ increases the required propellant mass because the acceleration of the servicer decreases for a given propulsion system of fixed thrust, resulting in a longer transfer. 
As such, similarly to the effect of $D$, an increase in $\msdry$ would move the design towards bringing the depots ``closer'' to the clients. 
This effect is most pronounced for facilities servicing a larger number of clients; when the optimal solution uses 6 depots, the facilities located at $\Omega = 30$, $150$, and $270$ \SI{}{deg} servicing 16, 14, and 13 clients respectively tend to reside at higher $a$ and lower $e$ comparing the solutions from $\msdry = 500$ \SI{}{kg} in Table \ref{tab:used_elts_FLP_gpsgal_ms500} to the solutions from $\msdry = 1,000$ \SI{}{kg} in Table \ref{tab:used_elts_FLP_gpsgal_ms500}.

\subsection{Depots Location Refinements}
Fixing the number of depots and the allocation of clients to those obtained from the OFLP solution, the depots' locations are refined in continuous space using metaheuristics. The refinement decision vector consists of the four orbital elements that have been discretized for OFLP, $\aleph = [a,e,i,\Omega]$, and this step is performed for each depot separately. 
For this demonstration, the facility locations for the three OFLP results with $D=1$ and $\msdry = 500\,\mathrm{kg}$ are refined. 
The optimization is done using Differential Evolution \cite{Price2013}, with a population size of 50 with a mutation scale factor $F = 0.9$. 

\begin{figure}
    \centering
    \includegraphics[width=0.95\linewidth]{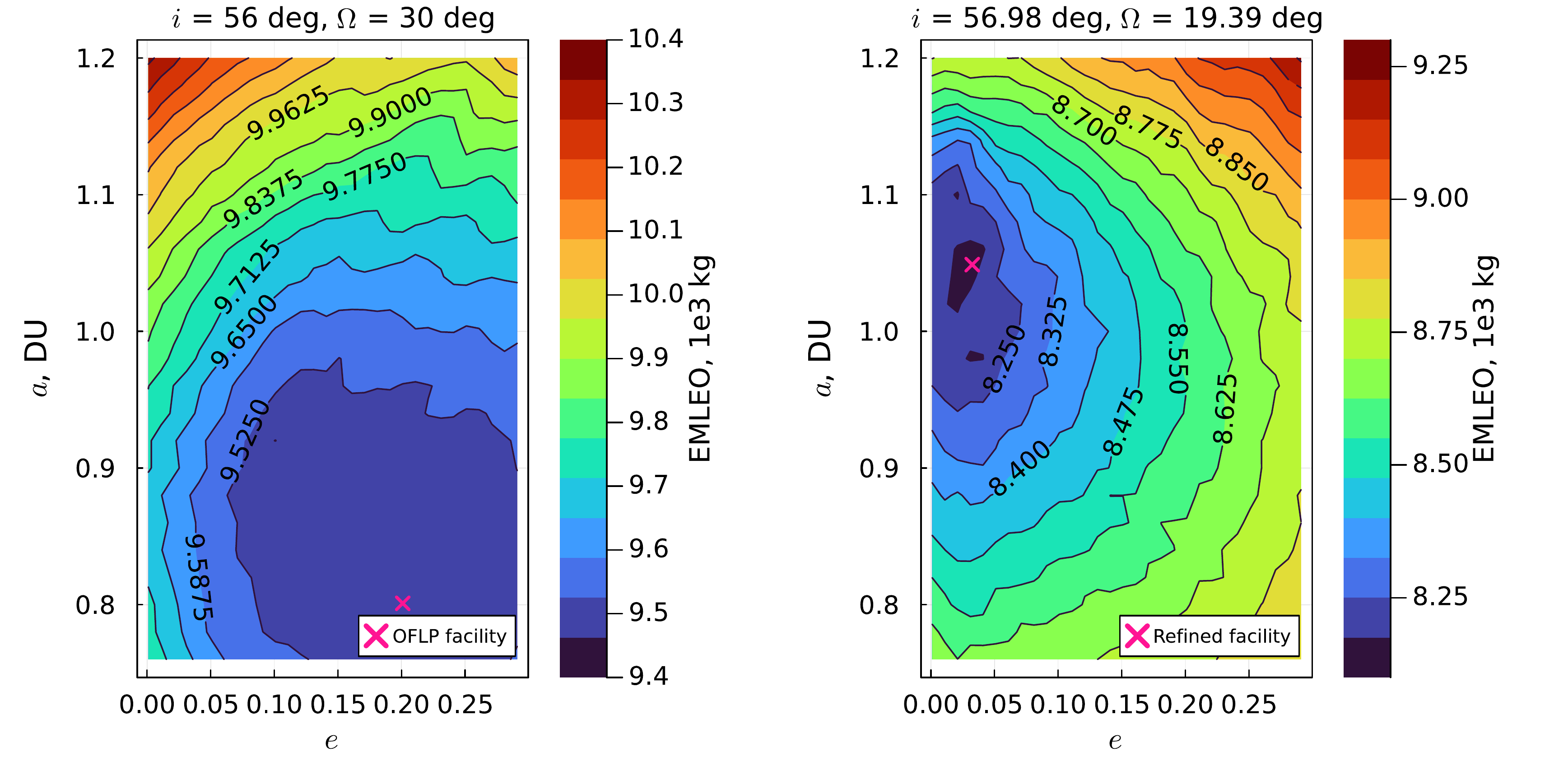}
    \caption{EMLEO contour of facility 1 for $\mddry = 1,500$ \SI{}{kg}, $\msdry = 500$ \SI{}{kg}, and $D = 1$, with inclination and RAAN fixed to the solution of OFLP (left) and inclination and RAAN fixed to the refined solution (right)}
    \label{fig:refined_contour_lvAriane64_dem1_dry1500}
\end{figure}

Table \ref{tab:refined_gnss_locations_D1} shows the refined locations of the facilities for $\msdry = 500$ \SI{}{kg} and $D = 1$. The refined results include both facilities that have been refined within the grid of the OFLP and ones that have moved outside in terms of $a$ and $e$. 
On the other hand, refinements in $i$ have mostly been within the grid except for facility 1 in the $\mddry = 1,500\,\mathrm{kg}$ and $\mddry = 2,000\,\mathrm{kg}$ cases, and the refinements in $\Omega$ have strictly been within the grid, to optimally align each facility with the cluster of clients. 
The occasional fluctuation in $a$, $e$, and $i$ beyond the grid is an artifact of the discretization, where previously in OFLP the transfer costs have been traded-off up to the precision limited by the coarseness of the grid as shown in Figure \ref{fig:contour_gnss}. 
The resulting reduction in total EMLEO is within 7\% for all tested cases. For individual depots, the largest reductions are observed for facility 1 in $\mddry = 1,500\,\mathrm{kg}$ and $\mddry = 2,000\,\mathrm{kg}$, where the reduction is within 15\%. 
To understand this fluctuation, taking this facility 1 as an example, Figure \ref{fig:refined_contour_lvAriane64_dem1_dry1500} shows the contour of the objective \eqref{eq:refinement_nlp_objective} in semimajor axis and eccentricity space, while freezing the inclination and RAAN to the values found by the OFLP or by the refinement NLP, respectively. 
It is possible to observe that the local minimum of EMLEO clearly shifts as the inclination and RAAN are adjusted, resulting in a noticeably different facility orbit. Overall, the results demonstrate the value of the refinement step by further improving the OFLP results.

Note that these refined solutions are assuming a fixed servicing allocation based on the OFLP results, and so could not have been obtained without first formulating and solving an OFLP instance. Specifically, concurrently optimizing the number of facilities, their locations, and the allocation of each client to each specific facility altogether is prohibitively difficult in a traditional nonlinear global optimization method because it would require large numbers of both binary and continuous variables, which would severely degrade the performance of any global optimization algorithm. 
In contrast, the proposed scheme combining the OFLP and the continuous refinement problem leverages the strength of both formulations, enabling the design of complex OSAM depot placements. 

\begin{figure}
    \centering
    \includegraphics[width=0.8\linewidth]{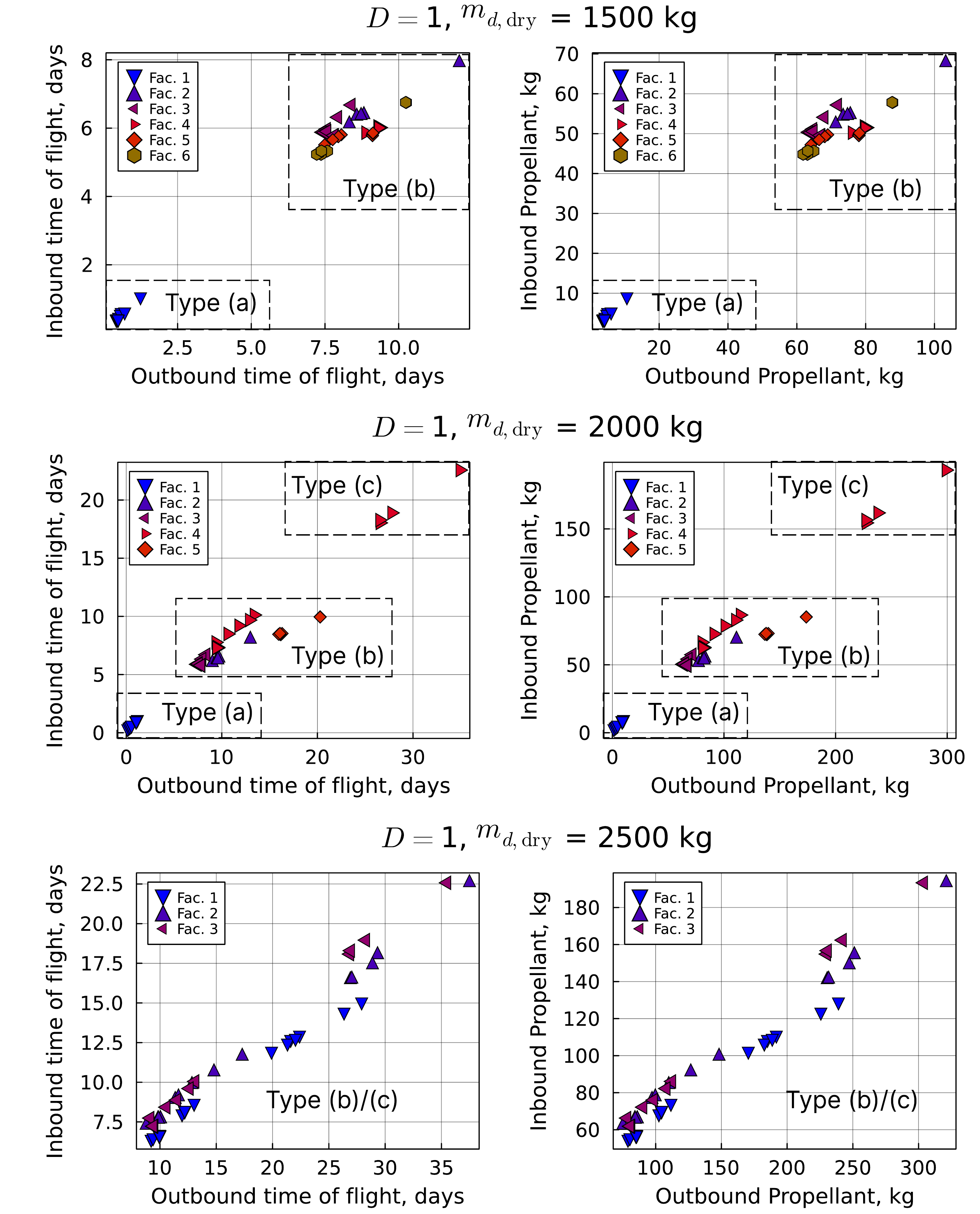}
    \caption{Transfer time and propellant mass of outbound and inbound trips used by refined solutions with $\msdry = 500$ \SI{}{kg} and $D = 1$}
    \label{fig:refined_prop_mass_scatter}
\end{figure}

\begin{figure}
    \centering
    \includegraphics[width=0.95\linewidth]{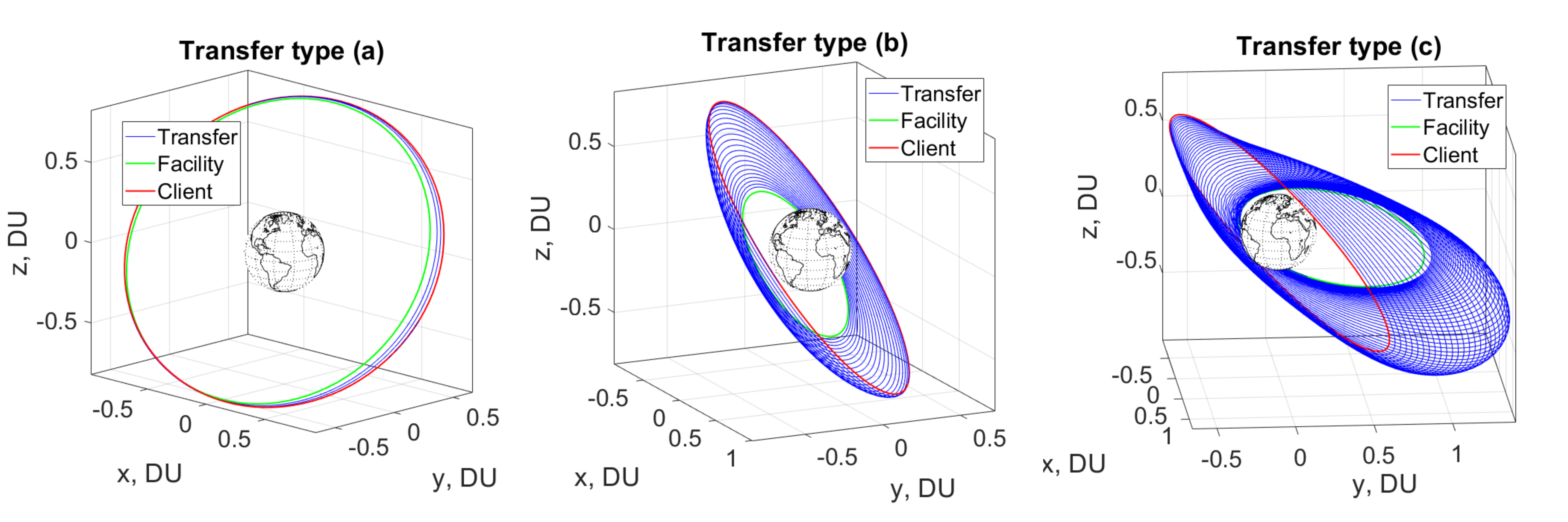}
    \caption{Examples of the three transfer types observed in the refined solutions with $\msdry = 500$ \SI{}{kg} and $D = 1$}
    \label{fig:transfer_types}
\end{figure}

For the refined depot location solutions, Fig. \ref{fig:refined_prop_mass_scatter} shows the time of flight and propellant mass of the outbound and inbound transfers. As previously highlighted, the outbound trip requires a longer time and more propellant due to the additional mass to be carried. Also, the time of flight and propellant mass consumption are proportional, as the transfers are assumed to use maximum thrust at all times. 
Across the three cases, there are roughly three groups of transfers that can be identified: 
\begin{enumerate}[label=(\alph*)]
    \item The first type is the cheapest and fastest, with a time of flight of around 2 days, with propellant consumption of less than 20 \SI{}{kg}; this corresponds to transfers from facilities with a relatively large semimajor axis and low eccentricity servicing clients in close-by orbital planes, such as facility 1 for $D=1$ with $\mddry = 1,500$ \SI{}{kg} and   $\mddry = 2,000$ \SI{}{kg}. 
    \item The second type of transfer has a time of flight of around 7 to 20 days, and a propellant consumption of around $50$ to $100$ \SI{}{kg}; this corresponds to transfers from facilities with a relatively low semimajor axis and high eccentricity, also servicing clients in similar orbital planes.
    \item The third type of transfer has higher times of flight as well as propellant consumption than the previous two types; this corresponds to transfers to clients in significantly different RAAN. 
\end{enumerate}
Figure \ref{fig:transfer_types} shows examples of each of these types of transfer for the outbound leg. 
Type (a) has a time of flight of \SI{1.17}{days} with a propellant consumption of \SI{9.99}{kg}, type (b) has a time of flight of \SI{7.93}{days} with a propellant consumption of \SI{67.95}{kg}, and type (c) has a time of flight of \SI{26.86}{days} with a propellant consumption of \SI{230.06}{kg}. 

\begin{table}[]
    \centering
    \caption{Summary of refined GPS \& Galileo servicing depots with $\msdry = 500$ \SI{}{kg} and $D = 1$}
    \begin{tabular}{llllllllll}
    \hline
    \hline
    \begin{tabular}[c]{@{}l@{}} $\mddry$ \\ \SI{}{kg} \end{tabular} &
    No. &
    $a$, \SI{}{DU}   & $e$   & $i$, \SI{}{deg}  & $\Omega$, \SI{}{deg}  & 
    \begin{tabular}[c]{@{}l@{}} $\mdwet$, \SI{}{kg} \\ (\% change) \end{tabular} &
    \begin{tabular}[c]{@{}l@{}} Depot EMLEO, \SI{}{kg} \\ (\% change) \end{tabular} &
    \begin{tabular}[c]{@{}l@{}} Total EMLEO, \SI{}{kg} \\ (\% change) \end{tabular}
    \\
    \hline
    \multirow{6}{*}{1,500}
    & 1 & 1.0488 & 0.0325 & 56.98 & 19.39 & 5,037  (-19.41\%) & 8,136  (-14.09\%) & \multirow{6}{*}{35,621 (-6.90\%)}\\
    & 2 & 0.5606 & 0.5375 & 54.62 & 90.59 & 2,728  (-1.10\%) & 4,187  (-2.89\%)  \\
    & 3 & 0.5812 & 0.5661 & 54.67 & 139.74 & 4,571  (-6.01\%) & 7,106  (-6.19\%)  \\
    & 4 & 0.5110 & 0.4936 & 54.56 & 202.21 & 2,461  (-1.62\%) & 3,663  (-3.59\%)  \\
    & 5 & 0.5571 & 0.5009 & 54.91 & 260.28 & 4,579  (-2.24\%) & 6,976  (-4.69\%)  \\
    & 6 & 0.6122 & 0.3192 & 54.72 & 324.97 & 3,676  (-0.31\%) & 5,552  (-0.90\%)  \\
    \hline
    \multirow{5}{*}{2,000}
    & 1 & 0.9782 & 0.0391 & 56.42 & 19.54 & 5,883  (-14.45\%) & 9,376  (-11.09\%) & \multirow{5}{*}{40,397 (-4.70\%)}\\
    & 2 & 0.5382 & 0.5187 & 54.79 & 90.62 & 3,278  (-13.67\%) & 4,966  (-1.10\%)  \\
    & 3 & 0.5813 & 0.5661 & 54.77 & 139.79 & 5,074  (-5.64\%) & 7,889  (-5.85\%)  \\
    & 4 & 0.6218 & 0.5829 & 51.84 & 243.28 & 7,501  (+0.38\%) & 11,871  (-0.86\%)  \\
    & 5 & 0.4831 & 0.4770 & 54.18 & 325.73 & 4,306  (-0.74\%) & 6,295  (-1.24\%)  \\
    \hline
    \multirow{3}{*}{2,500} 
    & 1 & 0.8762 & 0.5199 & 52.68 & 1.99 & 10,781  (+0.53\%) & 18,116  (-0.31\%) & \multirow{3}{*}{45,008 (-0.62\%)}\\
    & 2 & 0.6285 & 0.5871 & 52.42 & 120.60 & 8,967  (+0.27\%) & 14,232  (-0.67\%)  \\
    & 3 & 0.6204 & 0.5818 & 51.51 & 243.56 & 8,004  (+0.30\%) & 12,659  (-0.99\%)  \\
    \hline
    \hline
    \end{tabular}
    \label{tab:refined_gnss_locations_D1}
\end{table}

Noting the GPS Block III launch mass of \SI{3,880}{kg} or the Galileo launch mass of \SI{700}{kg} the mass of the depots are of similar scales to a single GPS satellite, and up to a few times heavier than a single Galileo satellite. 
However, a single facility is able to service and hence extend the lifetime of multiple satellites, hence reducing the overall cost of the programs.  
Taking for example the servicing architecture with $\mddry= 1,500\,\mathrm{kg}$ and $D=1$, the total architecture mass is $23,052$ \SI{}{kg}, while replacing all 31 GPS satellites and all 28 Galileo satellites would require launching $139,880$ \SI{}{kg}.

\subsection{A Posteriori Analysis with Multiclient trips}
The effect of conducting multiclient servicing trips is considered based on the allocation from the OFLP and the refined facility locations. Specifically, we focus the analysis on:
\begin{enumerate}
    \item depot number 1 from $\msdry = 500 \,\mathrm{kg}$ and $D=1$ with $\mddry = 1500 \,\mathrm{kg}$, with orbital elements $[a,e,i,\Omega] = [1.0488 \,\mathrm{DU}, 0.0325, 56.98^{\circ}, 19.39^{\circ}]$ allocated to 16 clients,    
    \item depot number 5 from $\msdry = 500 \,\mathrm{kg}$ and $D=1$ with $\mddry = 1500 \,\mathrm{kg}$, with orbital elements $[a,e,i,\Omega] = [0.5571 \,\mathrm{DU}, 0.5009, 54.91^{\circ}, 260.28^{\circ}]$ allocated to 13 clients, and   
    \item depot number 1 from $\msdry = 500 \,\mathrm{kg}$ and $D=1$ with $\mddry = 2500 \,\mathrm{kg}$, with orbital elements $[a,e,i,\Omega] = [0.8762 \,\mathrm{DU}, 0.5199, 52.68^{\circ}, 1.99^{\circ}]$ allocated to 23 clients. 
\end{enumerate}
All three depots are allocated to both GPS and Galileo satellites; the first two cases are representative examples of depots that are located on similar orbital planes as its allocated client satellites, with near-circular or elliptical facility orbits; the third case is a representative example of depots placed in an orbital plane between two orbital planes of its allocated client satellites, as shown for example on the bottom left window in Figure \ref{fig:pcomb_gnss_euus_prob201_loose}. 

\begin{figure}
     \centering
     \begin{subfigure}[b]{0.92\linewidth}
         \centering
         \includegraphics[width=\textwidth]{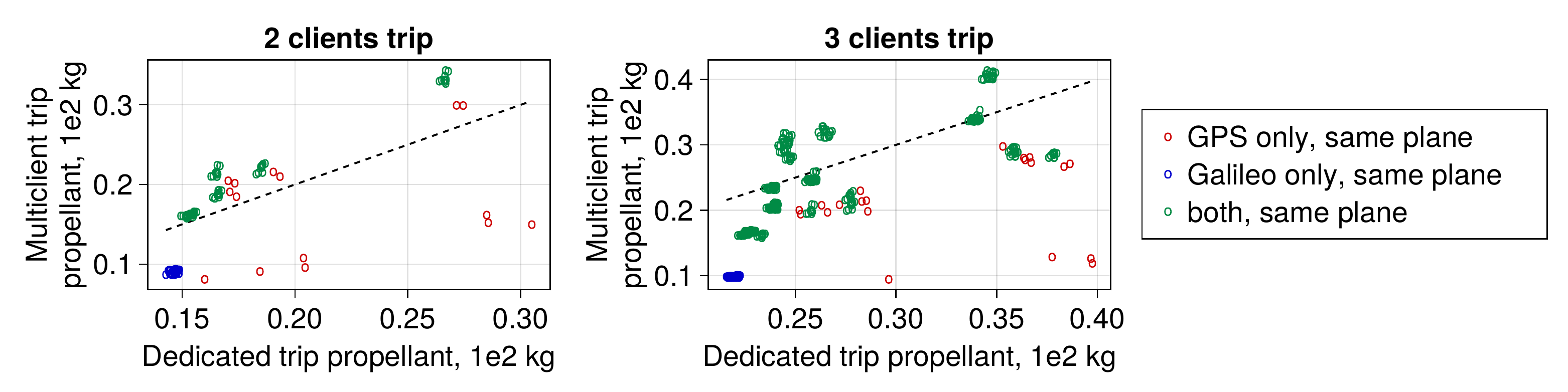}
         \caption{Depot 1 for $\msdry = 500$ \SI{}{kg}, $D=1$, and $\mddry = 1,500$ \SI{}{kg}, at $[a,e,i,\Omega] = [1.0488 \,\mathrm{DU}, 0.0325, 56.98^{\circ}, 19.39^{\circ}]$}
         \label{fig:multitrip_gnss_euus_prob201_dem1_dry1500_fac6}
     \end{subfigure}
     \begin{subfigure}[b]{0.92\linewidth}
         \centering
         \includegraphics[width=\textwidth]{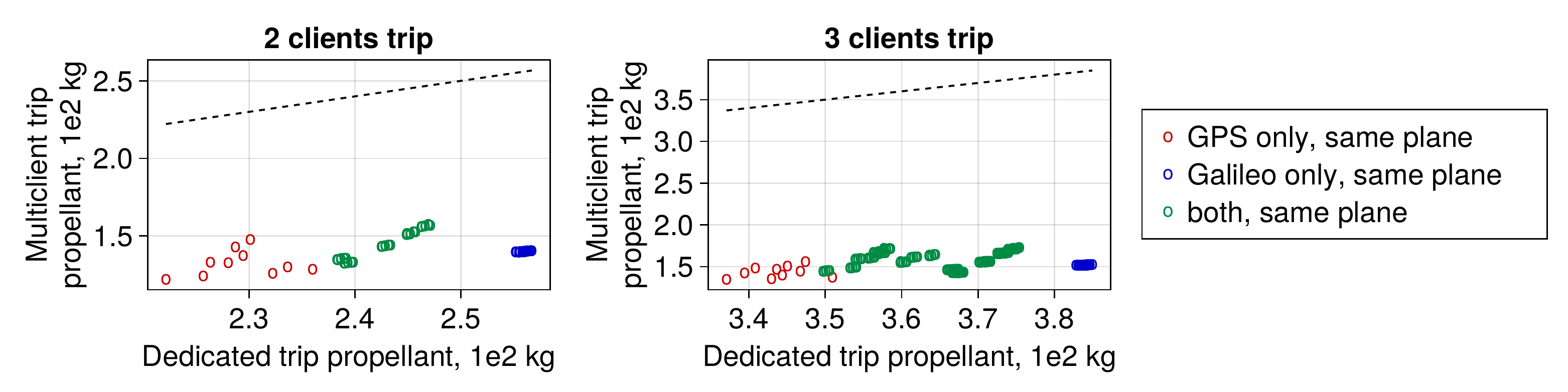}
         \caption{Depot 5 for $\msdry = 500$ \SI{}{kg}, $D=1$, and $\mddry = 1,500$ \SI{}{kg}, at $[a,e,i,\Omega] = [0.5571 \,\mathrm{DU}, 0.5009, 54.91^{\circ}, 260.28^{\circ}]$}
         \label{fig:multitrip_gnss_euus_prob201_dem1_dry1500_fac5}
     \end{subfigure}
     \begin{subfigure}[b]{0.92\linewidth}
         \centering
         \includegraphics[width=\textwidth]{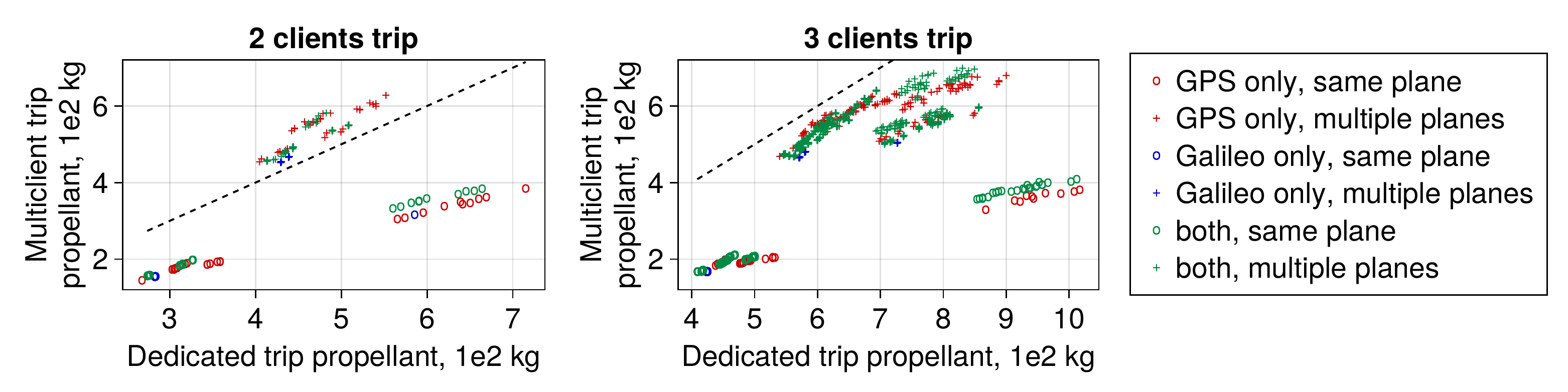}
         \caption{Depot 1 for $\msdry = 500$ \SI{}{kg}, $D=1$, and $\mddry = 2,500$ \SI{}{kg}, at $[a,e,i,\Omega] = [0.8762 \,\mathrm{DU}, 0.5199, 52.68^{\circ}, 1.99^{\circ}]$}
         \label{fig:multitrip_gnss_euus_prob201_dem1_dry2500_fac3}
     \end{subfigure}
        \caption{
        Propellant consumption comparison between a multiclient trip and the sum of dedicated trips, computed for all combinations of 2 (left column) and 3 (right column) clients allocated to each depot
        }
        \label{fig:multitrip_gnss_euus_prob201_dem1}
\end{figure}

We limit the analysis to up to $3$ client satellites to be serviced in a single trip, keeping in mind the aforementioned infrequent technology refresh rate for MEO constellations.
For each selected facility, the optimal cost \eqref{eq:multiclient_cost_optimal} is computed for all combinations of two or three of the allocated client satellites. 
The trade-off between conducting multiclient trips or separate, dedicated trips comes down to whether it is worth carrying the additional payload and propellant necessary to visit multiple clients instead of returning to the depot after each service. 
In Figure \ref{fig:multitrip_gnss_euus_prob201_dem1}, each data point corresponds to the propellant mass used from performing a multiclient trip \eqref{eq:multiclient_cost_optimal} to a set of two (left column) or three (right column) clients against the sum of the total propellant masses used to perform single client trips to each of these clients. The dotted lines denote the break-even cost, above which conducting dedicated trips is more propellant-effective, and below which conducting a multiclient trip is more effective. 

Firstly, for case 1) shown in Figure \ref{fig:multitrip_gnss_euus_prob201_dem1_dry1500_fac6}, the efficiency of multiclient trips depends on the clients that are bundled. For multiclient trips involving Galileo satellites only, multiclient trips are found to be always more efficient; this can be explained by the separation in the semimajor axis between the Galileo satellites (at $a \approx 1.114$ \SI{}{DU}) from the depot; hence, bundling the orbit-raising effort into a single trip reduces the total propellant necessary for servicing. 
In contrast, since the GPS satellites are at $a \approx 1.0$ \SI{}{DU}, the benefit of bundling the clients is diminished. 
The fact that the depot resides at a semimajor axis between the two constellations makes a 2-client multiclient trip to both GPS and Galileo clients inefficient, as visible from the green markers, as the servicer effectively ``passes through'' the depot's orbit anyway during the transit between the clients. Meanwhile, for 3-client multiclient trips, there is a trade-off depending on whether the 3 clients consist of 2 GPS satellites or 2 Galileo satellites. 

Secondly, from Figure \ref{fig:multitrip_gnss_euus_prob201_dem1_dry1500_fac5}, both 2- and 3-client multiclient trips are consistently more efficient. 
Due to the construction of the objective in the OFLP \eqref{eq:oflp_prob_objective} as well as in the refinement problem \eqref{eq:refinement_nlp_objective}, the optimal solution strikes a balance between the effort necessary to launch a given mass into the depot's orbit and the round-trip costs between the depot and its allocated clients; two viable strategies are to either launch a larger $\mathrm{EMLEO}$ into orbit such that the depot is placed closer to the constellation(s) it services, as is the case for the depot from case 1) shown in Figure \ref{fig:multitrip_gnss_euus_prob201_dem1_dry1500_fac6}, or launch a smaller $\mathrm{EMLEO}$ into a depot orbit that is further away from its allocated constellation(s), as is the case for the depot from case 2) shown in Figure \ref{fig:multitrip_gnss_euus_prob201_dem1_dry1500_fac5}. 
If the latter strategy is adopted, multiclient trips may lead to propellant savings, as exemplified by the studied case shown in Figure \ref{fig:multitrip_gnss_euus_prob201_dem1_dry1500_fac5}: for the specific servicer and servicing hyperparameters chosen, the significantly large maneuver to bring the servicer from the depot's orbit to the clients' has a stronger dependency on the number of round trips the servicer must do than on the mass it has to carry through. 
This effect is more pronounced for visiting Galileo clients than GPS clients since the former constellation is at a higher altitude, making them even more distant from the depot. 
Thus, when both strategies provide comparative performance, the expected servicing frequency may play a decisive role in deciding which type of depot orbit to employ. 

Finally, for case 3) depicted in Figure \ref{fig:multitrip_gnss_euus_prob201_dem1_dry2500_fac3}, all 2-client multiclient trips with clients residing on a single orbital plane are found to be more efficient than their dedicated trip counterparts, while all 2-client multiclient trips with one client on each orbital plane are found to be less efficient; in the 3-client multiclient trips, since at least 2 clients are collocated on the same orbital plane, multiclient trips are found to be more efficient in all cases. 
When 2 clients are located on the same orbital plane, the dominant cost arising from the orbital plane change to be conducted by the servicer dictates the efficacy of bundling the servicing; this is akin to case 2), where bundling the maneuver to adjust the semimajor axis and eccentricity from the depot's to the clients' was found to be advantageous. 
Interestingly, even when the clients are on two different orbital planes, the 2-client multiclient trips do not incur a significant increase in propellant cost compared to their dedicated trip counterparts. 
This can be attributed to the fact that orbital plane change cost is much lower at higher radii from the Earth; hence, the propellant cost for traversing from one client orbital plane to the other is significantly reduced when the servicer does not need to lower its semimajor axis down to the depot's in between.

\section{Conclusion}
\label{sec:conclusion}
This work studies the optimal placement of on-orbit servicing depots for satellite constellations at MEO or higher altitudes. The number of servicing depots as well as their placement(s), from which servicing vehicles fly to rendez-vous with the client satellites, are simultaneously optimized by modifying the facility location problem to formulate the Orbital Facility Location Problem (OFLP). 
Potential locations for placing the depots are considered in a discretized design space consisting of orbital slots. 
The combinatorial nature of the orbital slots results in large numbers of transfers to be computed; this is done by using the Q-law backward in time to obtain the return trip propellant mass expenditure. 
The cost associated with assigning a facility on a particular orbital slot to a client satellite as well as the cost of establishing this facility is combined to form a single objective based on the sum of each facility's EMLEO. 
Once a solution for the OFLP has been obtained, the facility locations have been refined using a continuous global optimization scheme while keeping the allocations of clients to each facility fixed to that of the OFLP solution. 

The resulting OFLP still retains the form of a binary linear program, which can be solved efficiently even with a large number of variables. 
The formulation is applied to a scenario for placing depots to service the GPS constellation, Galileo constellation, and the two constellations simultaneously.  
Multiple instances of the OFLP have been solved with different values for the depot and servicer parameters, and the significant variation in number, as well as the locations of the depots, have been observed. 
Intuitive design features, such as allocating clients in similar orbital planes to a single depot, have been confirmed to be advantageous by the OFLP, while less intuitive results, such as placing the depot at a slightly lower inclination than the clients to reduce the propellant cost of traversing in RAAN, have also been discovered. 
An a posteriori analysis based on servicing trips visiting multiple clients has revealed supplementary insights on the type of depot orbits as well as their associated servicing trips; it has been found that servicing trips visiting multiple clients in a single round trip is particularly advantageous for depots located on elliptical orbits, both in the case of depots allocated to clients on a single orbital plane as well as depots located to allocated clients residing on two separate orbital planes. 

The methods presented in the work, along with the analyses based on the GPS and Galileo satellites, enable studying the feasibility of on-orbit servicing architectures of high-altitude satellite constellations. 
As on-orbit servicing technology matures in the GEO market, the MEO and GSO markets would present a natural extension of such service and high-level design of such architecture may be conducted with the OFLP. 

\pagebreak
\section*{Appendix}
\subsection{Expressions for Q-Law}
\subsubsection{Maximum rates of change of orbital element $\dot{\elements}_{x x}$}
\begin{align}
    \dot{a}_{x x} &=2 F a \sqrt{\frac{a}{\mu}} \sqrt{\frac{1+\sqrt{f^{2}+g^{2}}}{1-\sqrt{f^{2}+g^{2}}}} \\
    \dot{f}_{x x} & \approx 2 F \sqrt{\frac{p}{\mu}} \\
    \dot{g}_{x x} & \approx 2 F \sqrt{\frac{p}{\mu}} \\
    \dot{h}_{x x} &=\frac{1}{2} F \sqrt{\frac{p}{\mu}} \frac{s^{2}}{\sqrt{1-g^{2}}+f} \\
    \dot{k}_{x x} &=\frac{1}{2} F \sqrt{\frac{p}{\mu}} \frac{s^{2}}{\sqrt{1-f^{2}}+g}
\end{align}
where $F$ is the magnitude of the perturbing force. 

\subsection{Keplerian Elements of Considered Constellations}
Tables \ref{tab:gps_clients} and \ref{tab:galileo_clients} show the Keplerian elements of the GPS and Galileo constellation fleet, respectively. 
\begin{table}[h]
\centering
\caption{Keplerian Elements of GPS Constellation}
\begin{tabular}{@{}lrrrrr@{}}
\toprule
Satellite number & \multicolumn{1}{l}{$a$, \SI{}{km}} & \multicolumn{1}{l}{$e$} & \multicolumn{1}{l}{$i$, \SI{}{deg}} & \multicolumn{1}{l}{$\Omega$, \SI{}{deg}} & \multicolumn{1}{l}{$\omega$, \SI{}{deg}} \\ \midrule
 1 & 26560.355 & 6.4584e-03 & 55.53 & 150.07 & 53.20 \\
 2 & 26560.460 & 4.7800e-03 & 54.18 & 72.93 & 188.43 \\
 3 & 26561.192 & 1.3721e-02 & 55.12 & 146.99 & 254.56 \\
 4 & 26561.008 & 1.2823e-02 & 55.42 & 267.35 & 41.45 \\
 5 & 26560.439 & 2.4678e-02 & 55.07 & 17.50 & 309.60 \\
 6 & 26560.919 & 8.8526e-03 & 55.91 & 328.36 & 127.48 \\
 7 & 26572.909 & 2.0378e-02 & 55.39 & 17.68 & 280.51 \\
 8 & 26560.094 & 1.4080e-02 & 55.97 & 325.81 & 276.13 \\
 9 & 26559.723 & 1.0584e-02 & 54.70 & 203.57 & 25.15 \\
10 & 26560.771 & 8.3765e-03 & 55.43 & 266.30 & 75.05 \\
11 & 26560.023 & 1.4504e-02 & 53.36 & 134.59 & 65.84 \\
12 & 26559.797 & 2.0014e-03 & 56.10 & 326.58 & 147.45 \\
13 & 26559.858 & 1.6622e-02 & 54.46 & 202.48 & 232.26 \\
14 & 26559.181 & 5.7239e-03 & 55.19 & 79.74 & 64.40 \\
15 & 26559.538 & 1.0562e-02 & 54.72 & 261.70 & 56.85 \\
16 & 26560.119 & 1.1835e-02 & 56.66 & 23.12 & 53.36 \\
17 & 26560.353 & 1.3191e-02 & 53.52 & 197.47 & 48.46 \\
18 & 26560.199 & 1.0589e-02 & 55.60 & 322.15 & 38.13 \\
19 & 26560.529 & 6.0448e-03 & 53.61 & 203.00 & 209.29 \\
20 & 26559.042 & 2.7915e-03 & 56.62 & 22.64 & 304.45 \\
21 & 26560.934 & 2.4195e-03 & 54.72 & 141.08 & 112.11 \\
22 & 26561.428 & 4.2503e-03 & 55.93 & 82.28 & 58.76 \\
23 & 26559.493 & 7.3252e-03 & 53.62 & 258.80 & 20.82 \\
24 & 26559.720 & 7.7127e-03 & 55.09 & 320.89 & 7.69 \\
25 & 26560.447 & 8.2834e-03 & 55.92 & 82.17 & 217.98 \\
26 & 26560.306 & 6.3642e-03 & 54.94 & 141.77 & 229.63 \\
27 & 26561.026 & 2.1567e-03 & 55.12 & 144.23 & 187.36 \\
28 & 26560.756 & 2.9509e-03 & 55.74 & 23.41 & 183.78 \\
29 & 26559.913 & 2.8304e-03 & 55.64 & 80.65 & 180.09 \\
30 & 26560.207 & 2.4235e-03 & 54.48 & 264.26 & 191.16 \\
31 & 26560.209 & 8.7880e-04 & 55.25 & 25.27 & 207.44 \\
\bottomrule
\end{tabular}
\label{tab:gps_clients}
\end{table}

\begin{table}[h]
\centering
\caption{Keplerian Elements of Galileo Constellation}
\begin{tabular}{@{}lrrrrr@{}}
\toprule
Satellite number & \multicolumn{1}{l}{$a$, \SI{}{km}} & \multicolumn{1}{l}{$e$} & \multicolumn{1}{l}{$i$, \SI{}{deg}} & \multicolumn{1}{l}{$\Omega$, \SI{}{deg}} & \multicolumn{1}{l}{$\omega$, \SI{}{deg}} \\ \midrule
 1 & 29600.198 & 4.8800e-05 & 57.04 & 17.43 & 2.09 \\
 2 & 29600.168 & 2.7400e-04 & 57.04 & 17.46 & 298.10 \\
 3 & 29600.354 & 1.8980e-04 & 55.18 & 137.75 & 231.25 \\
 4 & 29600.327 & 1.5750e-04 & 55.18 & 137.80 & 138.07 \\
 5 & 27977.498 & 1.6173e-01 & 50.18 & 326.56 & 129.27 \\
 6 & 27977.430 & 1.6145e-01 & 50.21 & 325.63 & 130.07 \\
 7 & 29600.058 & 4.4700e-04 & 56.84 & 17.45 & 239.84 \\
 8 & 29600.127 & 4.4590e-04 & 56.84 & 17.53 & 226.09 \\
 9 & 29600.445 & 5.0140e-04 & 55.54 & 258.02 & 2.52 \\
10 & 29600.444 & 3.5460e-04 & 55.54 & 258.04 & 345.16 \\
11 & 29600.271 & 1.8730e-04 & 55.20 & 137.45 & 344.51 \\
12 & 29600.294 & 1.3840e-04 & 55.19 & 137.52 & 1.47 \\
13 & 29600.440 & 4.7190e-04 & 55.68 & 257.99 & 324.87 \\
14 & 29600.265 & 4.0670e-04 & 55.68 & 257.97 & 296.90 \\
15 & 29600.433 & 1.5940e-04 & 54.86 & 137.62 & 310.57 \\
16 & 29600.438 & 1.6710e-04 & 54.85 & 137.64 & 29.68 \\
17 & 29600.412 & 1.6100e-04 & 54.86 & 137.60 & 248.61 \\
18 & 29600.434 & 7.4200e-05 & 54.85 & 137.64 & 230.50 \\
19 & 29600.355 & 3.4110e-04 & 55.66 & 257.78 & 293.86 \\
20 & 29600.347 & 5.1140e-04 & 55.66 & 257.84 & 290.54 \\
21 & 29600.349 & 4.7900e-04 & 55.66 & 257.82 & 302.25 \\
22 & 29600.344 & 5.4600e-04 & 55.66 & 257.79 & 285.85 \\
23 & 29600.073 & 4.5590e-04 & 57.22 & 17.33 & 259.06 \\
24 & 29600.076 & 3.9160e-04 & 57.22 & 17.38 & 263.93 \\
25 & 29600.077 & 3.5470e-04 & 57.22 & 17.36 & 262.57 \\
26 & 29600.077 & 4.2110e-04 & 57.22 & 17.30 & 260.34 \\
27 & 29600.059 & 3.8470e-04 & 57.21 & 17.22 & 213.15 \\
28 & 29600.059 & 2.7530e-04 & 57.21 & 17.21 & 204.09 \\
\bottomrule
\end{tabular}
\label{tab:galileo_clients}
\end{table}

\pagebreak
\newpage
\bibliography{references}

\end{document}